\documentclass[prb,showpacs,twocolumn,superscriptaddress]{revtex4-1}
\usepackage{amsmath,graphicx}
\usepackage{amsfonts}

\newcount\minute
\newcount\hour
\newcount\hourMins
\def\now
{
  \minute=\time
  \hour=\time \divide \hour by 60
  \hourMins=\hour \multiply\hourMins by 60
  \advance\minute by -\hourMins
  \zeroPadTwo{\the\hour}:\zeroPadTwo{\the\minute}
}
\def\timestamp
{
  \today\ \now
}
\def\today
{
  \zeroPadTwo{\the\day}-\zeroPadTwo{\the\month}-\the\year%
}
\def\zeroPadTwo#1%
{
  \ifnum #1<10 0\fi
  #1%
}

\def \dif {{d}}
\def \Dif {\mathcal{D}}

\newcommand{\be}{\begin{equation}}
\newcommand{\ee}{\end{equation}}

\newcommand{\bea}{\begin{eqnarray}}
\newcommand{\eea}{\end{eqnarray}}

\newcommand{\nn}{\nonumber}

\date{\timestamp}
\begin{document}

\title{Exact form of the exponential correlation function in the glassy super-rough phase}

\author{Pierre Le Doussal}
\affiliation{Laboratoire de Physique Th\'{e}orique--CNRS, Ecole Normale Sup\'{e}rieure, 24 rue Lhomond, 75005 Paris, France}

\author{Zoran Ristivojevic}
\affiliation{Centre de Physique Th\'{e}orique, Ecole Polytechnique, CNRS, 91128 Palaiseau, France}
\affiliation{Laboratoire de Physique Th\'{e}orique--CNRS, Ecole Normale Sup\'{e}rieure, 24 rue Lhomond, 75005 Paris, France}

\author{Kay J\"org Wiese}
\affiliation{Laboratoire de Physique Th\'{e}orique--CNRS, Ecole Normale Sup\'{e}rieure, 24 rue Lhomond, 75005 Paris, France}

\begin{abstract}
We consider the random-phase sine-Gordon model in two dimensions. It describes two-dimensional   elastic systems with random periodic disorder, such as pinned flux-line arrays, random field XY models, and surfaces of disordered crystals. The model exhibits a super-rough glass phase at low temperature $T<T_{c}$ with relative displacements growing with distance $r$ as $\overline{\langle [\theta(r)-\theta(0)]^2\rangle} \simeq A(\tau) \ln^2 (r/a)$,
where $A(\tau) = 2 \tau^2- 2 \tau^3 +\mathcal{O}(\tau^4)$ near the transition and $\tau=1-T/T_{c}$. We calculate all higher cumulants and show that
they grow as  $\overline{\langle[\theta(r)-\theta(0)]^{2n}\rangle}_c \simeq [2 (-1)^{n+1} (2n)! \zeta(2n-1) \tau^2 + \mathcal{O}(\tau^3) ] \ln(r/a)$, $n \geq 2$, where $\zeta$ is the Riemann zeta function. By summation, we
obtain the decay of the exponential correlation function as $\overline{\langle e^{iq\left[\theta(r)-\theta(0)\right]}\rangle} \simeq (a/r)^{\eta(q)}
\exp\boldsymbol(-\frac{1}{2}\mathcal{A}(q)\ln^2(r/a)\boldsymbol)$ where $\eta(q)$ and ${\cal A}(q)$ are obtained for arbitrary $q \leq 1$ to leading order
in $\tau$. The anomalous exponent is  $\eta(q) = c q^2 - \tau^2 q^2 [2\gamma_E+\psi(q)+\psi(-q)]$
in terms of the digamma function $\psi$, where $c$ is non-universal and $\gamma_E$ is the Euler constant. The correlation function shows a faster decay at $q=1$, corresponding to fermion operators in the dual picture, which should be visible in Bragg scattering experiments.

\end{abstract}

\maketitle

\section{Introduction and main results}

The random-phase sine-Gordon model is the simplest model to describe the effect of quenched disorder
on a periodic elastic system, the so-called random periodic class. \cite{Cardy+82,Toner+90,Hwa+94,Carpentier+97, GiamarchiLeDoussalBookYoung,fisher89vortex}
It models a host of experimental systems in presence of substrate impurities, such as charge-density waves,
vortex lattices, \cite{Blatter+94,Nattermann+00} random-field XY models,
\cite{Cardy+82,Feldman2001,TarjusTissier2005,FEDORENKO} and smectic liquid crystals \cite{RADZ1}
in situations where topological defects are absent, or can be ignored, allowing for an elastic description.
In the simplest situation, the displacement fields from the perfect periodic position are encoded by a $2 \pi$ periodic one-component phase field $\theta(x)$. In a one-dimensional crystal of spacing $a$ with elastic displacements $u(x)$, the phase field is $\theta(x)=G_0 u(x)$ where $G_0=2 \pi/a$
is the smallest reciprocal lattice vector. In three dimensions, the random-phase sine-Gordon model was used to predict the existence of a Bragg-glass phase
which is a glass pinned by (weak) quenched disorder, but which also retains topological order (no dislocations)
and nearly perfect translational order \cite{Giamarchi+95PhysRevB.52.1242,Nattermann+00, LeDoussal2010Book1,DSFISHER}
called quasi-order. Diffraction experiments \cite{KleinJoumardBlanchardMarcusCubittGiamarchiLeDoussal2001} measure the
spatial correlations of the field $e^{i q \theta(x)}$ for $q =1$. In that case, the elastic description predicts a power-law decay of these correlations due to quenched disorder,
\cite{Giamarchi+95PhysRevB.52.1242,Nattermann+00,bogner+01}
characteristic of a quasi-ordered phase and leading to divergent Bragg peaks in the experiments. At stronger disorder, the presence of free topological defects is expected to lead to an exponential decay of these correlations with distance, as quasi-order is destroyed.

Quasi-order usually arises when the phase field deformations $\theta(x)-\theta(0)$ grow logarithmically with scale $x$, leading to power law
scaling for the exponential field $e^{i q \theta(x)}$. To probe deeper the properties of the phase field $\theta(x)$ in
a quasi-ordered phase it would be useful to predict, and to measure, these correlations for $q$ not necessarily an integer.
One example with $q=1/2$ would be a spin density wave, e.g., of XY symmetry $S=A e^{i \phi}$, submitted to time-reversal invariant disorder, such as random anisotropy,\cite{Feldman2001} i.e., which couples to $2 \phi\equiv \theta$. Arbitrary values of $q$ would allow to probe the probability distribution of the phase deformations $\theta(x)-\theta(0)$ and to characterize the multi-fractal properties of the exponential field. Here we
restrict to the case of a $d=2$ dimensional periodic system, leaving the study of $d>2$ for an upcoming work. \cite{FEDO-US}

In two-dimensional periodic systems with quenched disorder thermal fluctuations play a more important role than in three-dimensional ones. As was discovered in the
pioneering work of \citet{Cardy+82} they induce a phase transition at some critical temperature
$T_c$ to a high temperature phase where disorder is irrelevant. For $T<T_c$ a glass phase exists in this model
which has been investigated in a number of works. \cite{Cardy+82,Toner+90,Hwa+94,Carpentier+97,Ristivojevic+12, perret+12PhysRevLett.109.157205}
A very nice realization of this model in terms of crystal surfaces was described by \citet{Toner+90}. Since it allows in principle to measure the exponential correlation for any $q$, we now recall the basic phenomenology of surfaces.
Note that another interesting realization of the Cardy-Ostlund model was obtained recently in the context of a smectic with surface
disorder. \cite{RADZ2,RADZ21}

Perfect crystals are characterized by an ideal lattice. At high temperatures, the thermal motion of the atoms overwhelms the lattice potential, and  the crystal melts. In the present article we consider physical effects that occur at significantly lower temperatures. Consider the
atoms at the surface of a crystal. They can  more easily be displaced from their equilibrium positions by  thermal fluctuations, since they reside at the boundary between the crystal and, usually,  a fluid. They feel the periodic potential created by the bulk of the crystal, as well as a more uniform potential from the fluid. At low temperatures,  surface atoms are not displaced  significantly from their equilibrium positions determined by the bulk potential, the surface is flat, and the atoms are ideally arranged. At higher temperatures, the thermal motion of surface atoms becomes significant, and  they are randomly displaced, forming a rough fluctuating surface. The periodic potential is destroyed by thermal effects, a phenomenon  known as the roughening transition. \cite{Chaikin+,Nozieres}

On the other hand, real crystals always experience some kind of disorder that tends to diminish the infinite correlation length of the translational order of a perfect crystal. In such situations, atoms at the surface do not longer experience a perfect periodic potential, but rather a disordered one created by the bulk. At high temperatures the disordered potential is washed out by thermal fluctuations and thus is unimportant for the shape of the surface. The surface is rough. On the contrary, at  temperatures $T$ below a critical temperature $T_{c}$,  surface atoms follow the disorder potential, thereby forming a surface that is even rougher. This phase transition is known as the super-roughening transition. \cite{Toner+90}

Using surface-sensitive scattering experiments, one can probe the crystal surface. One directly measures the disorder and thermally averaged correlation function
\begin{equation}\label{1}
C(q,r):=\overline{\langle e^{iq\left[\theta(r)-\theta(0)\right]}\rangle}\ ,
\end{equation}
 where $\theta$ denotes the two-dimensional height field of the surface, in units of $G_0=2\pi/a$. The quantity $q$ denotes the component of the wave vector perpendicular to the surface.

 In the high-temperature rough phase, at $T>T_c$, the correlation function $C(q,r)$ decays at large distances as a power law  $C(q,r)\sim (a/r)^{\eta(q)}$, where $a$ is the lattice constant. The naive argument is that at high temperatures the fluctuations of the surface are effectively Gaussian, which in two dimensions always produce quasi-long-range order characterized by a power-law decay of $C(q,r)$
 and $\eta(q) \sim q^2$. However, despite being an irrelevant operator at high temperatures,
 the lattice potential still plays an important role and produces in fact
 a nontrivial result for $\eta(q)$.
 \begin{figure}
\includegraphics[width=0.9\columnwidth]{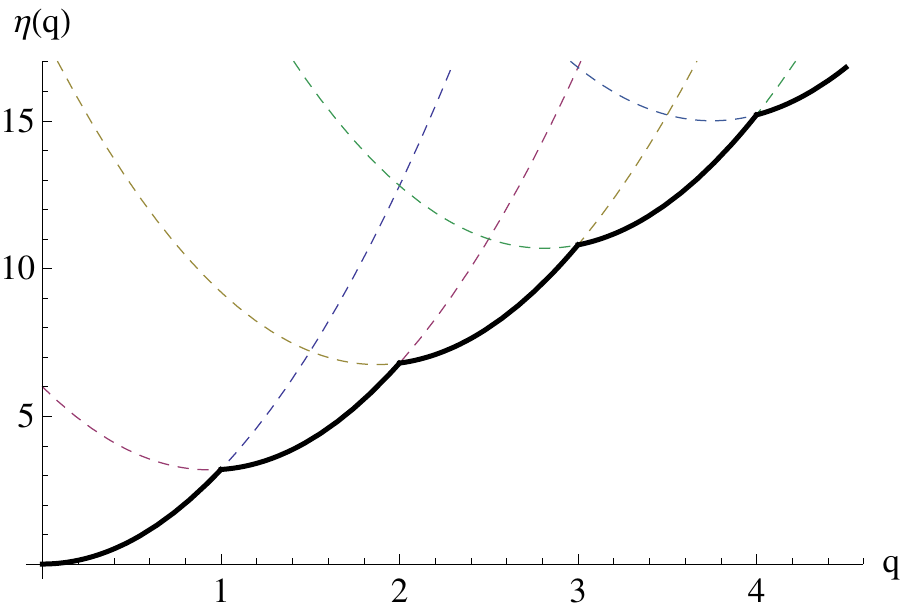}
\caption{The function $\eta(q)$ given in Eq.~(\ref{toner}) for $b=3,c=3.2$ is represented by the solid line. The functions $\eta_m(q)$  of Eq.~(\ref{toner1}) for $m=0,1,\ldots,4$ are given by the dashed lines.}\label{fig1}
\end{figure}
In their seminal paper, \citet{Toner+90} calculated $C(q,r)$ in the high-temperature rough phase, for $T>T_{c}$. They obtained the decay as a superposition of power laws $C(q,r) \simeq \sum_{m=1}^\infty A_{2 m} (a/r)^{\eta_m(q)}$ with exponents
\begin{eqnarray} \label{toner1}
\eta_m(q) &=& (c-b) q^{2} + b \left[ (q-m)^{2}+m \right].
\end{eqnarray}
At sufficiently large $r$ the smallest power dominates, hence they concluded
that the asymptotic exponent $\eta(q)$, for $q>0$, takes the form
\begin{eqnarray} \label{toner}
\eta(q) &=& \min_{m \in \mathbb{N}_{0}}  \eta_m(q) \\
 &\equiv& c q^2 + b \Big([q]^2 + [q] (1-2 q)\Big)\ , \nn
\end{eqnarray}
where $[q]$ is the integer part of $q$.   The coefficients of Eq.~(\ref{toner}) are
\begin{eqnarray}
c=\frac{2T}{T_{c}}\left(1+\frac{2T\sigma'}{T_{c}}\right), \quad b=\frac{2T}{T_{c}}.
\end{eqnarray}
They are given in terms of the critical temperature $T_{c}$ and a non-universal off-diagonal disorder $\sigma' \geq 0$  defined in Sec.~\ref{s:model}. While the coefficient $b$ is a universal function of the temperature, the coefficient $c$ depends on $\sigma'$ and hence is non-universal. (It depends on the details of the model at short scales, which can renormalize $\sigma'$). The form of Eqs.~(\ref{toner1}),(\ref{toner}) has a simple interpretation in the picture of the Coulomb gas put forward by \citet{Cardy+82} (see also below). In that picture, perturbation theory to order $2p$ in the disorder is equivalent to inserting $p$ pairs of replica $\pm 1$ charges. The vertex operator $e^{iq\left[\theta(r)-\theta(0)\right]}$ can itself be seen as inserting a charge $+q$ at $r$ and a charge $-q$ at $0$ in the same replica. For $q>1$, it is then energetically advantageous to {\em screen} the vertex operator $e^{iq\left[\theta(r)-\theta(0)\right]}$ with \(2m \) {\em replica charges} from the disorder leading to Eqs.~(\ref{toner1}) and (\ref{toner}). \footnote{Note that since it involves order $2m$ in perturbation
theory in the disorder, i.e., $A_{2m} \sim g^{2m}$, the result (\ref{toner}) holds only in the limit of $r \to \infty$ at fixed $q$. For a  large but fixed $r$, it will be more difficult to measure the behavior for larger $q$.}
As a consequence,  $\eta(q)$ always grows with $q, $  is  quadratic for  non-integer $q$, and continuous but with a cusp at integer values of $q$, see Fig.~\ref{fig1}.
An interesting possibility, suggested by the form of the perturbation theory of \citet{Toner+90} (i.e., the $A_{2m}$ appear to have alternating signs) is that
$C(q,r)$ is an oscillating function, changing sign at integer $q$. A simple toy model with such an oscillating behavior is given in Appendix
\ref{app:toy}. Finally, it is important to note that for
a {\it perfect crystal} (i.e., in the presence of commensuration only and with no disorder),
one has $\eta(q) = b (q- [1/2 + q])^2$, i.e., a periodic function of $q$ with cusps at {\it half integers} and minima at
all integers. \cite{Toner+90}

\begin{figure}
\includegraphics[width=0.9\columnwidth]{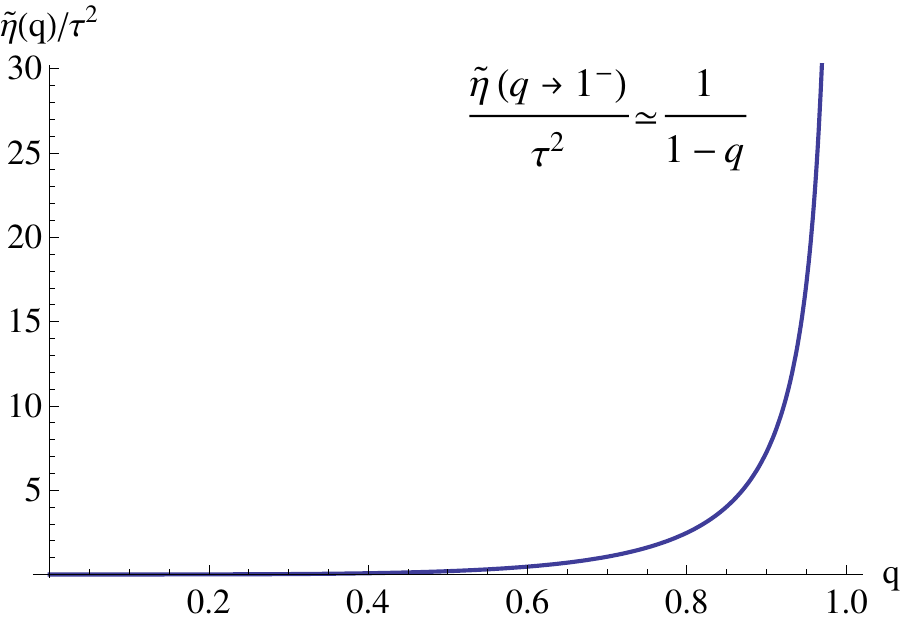}
\caption{Plot of the universal part of the anomalous exponent, $\widetilde\eta(q)$, for $0<q<1$.}\label{figetatilde}
\end{figure}

Let us now consider low temperatures, $T<T_{c}$. In this super-rough phase, the surface becomes even rougher \cite{Toner+90} producing a faster decay for $C(q,r)$ as a function of the distance $r$.
Although the general form for the decay of $C(q,r)$ was correctly anticipated in  Ref.~\onlinecite{bauer+96} in the context of
an $N$-component extension of the random-phase sine-Gordon model, its precise expression, including the value of the exponents,
has not yet been obtained.
The aim of the present article is to fill this gap and determine the exact form for $C(q,r)$ in the super-rough phase,
close to the super-roughening transition.
We will show that it takes the form
\begin{align}\label{G1form}
C(q,r)\simeq\left(\frac{a}{r}\right)^{\eta(q)}
\exp\left(-\frac{1}{2}\mathcal{A}(q)\ln^2(r/a)\right),
\end{align}
characterized by the anomalous exponent $\eta(q)$ and an amplitude $\mathcal{A}(q)$. We  obtained the exact dependence on $q$ of both quantities, to leading order in $\tau$, the distance from the transition, defined as
\begin{align}
\tau = 1 - T/T_{c}.
\end{align}
For $0< q<1$,  we find, up to terms of order $\mathcal{O}(\tau^3)$,
\begin{gather}\label{etafinal}
\eta(q) = c q^2 + \widetilde \eta(q), \\
\widetilde \eta(q) = -\tau^2 q^2\left[2\gamma_{{E}}+\psi (q)+\psi (-q)\right], \\
\label{Afinal}
\mathcal{A}(q) = 2q^2\tau^2.
\end{gather}
Here $\psi(q)$ is the digamma function.
By $\widetilde \eta(q)$ we denote the universal part of the exponent plotted in Fig.~\ref{figetatilde}; it  starts at $\mathcal{O}(q^4)$ at small $q$;  together
with $\mathcal{A}(q)$ they are  universal. The coefficient $c$ in $\eta(q)$ is non-universal, and within our cutoff scheme given by\begin{eqnarray}\label{ourc}
c = 2 (1-\tau)[1+2(1-\tau)\sigma']+\frac{\tau}{2} + \tau^2+\mathcal{O}(\tau^3),
\end{eqnarray}
where $\sigma'$ denotes a non-universal off-diagonal disorder defined below.

Both the amplitude and the exponent are functions of the temperature $T$, with $\mathcal{A}\equiv0$ for $T\ge T_{c}$;
$\eta(q)$ matches smoothly at $T_{c}$  to the expression (\ref{toner}), i.e.,
$\widetilde \eta(q)$ vanishes for $T \geq T_{c}$. An initial guess for the amplitude may be obtained from the variance of the height fluctuations at two points, which
we have recently calculated in Ref.~\onlinecite{Ristivojevic+12} to two loop accuracy,
\begin{gather}
\overline{\langle [\theta(r)-\theta(0)]^2\rangle} \simeq A(\tau) \ln^2 (r/a)+\mathcal{O}\boldsymbol(\ln(r/a)\boldsymbol),\\
A(\tau) = 2 \tau^2- 2 \tau^3 +\mathcal{O}(\tau^4).
\end{gather}
This result was  tested in a numerical simulation in Ref.~\onlinecite{perret+12PhysRevLett.109.157205}, using a dimer-model representation of the sine-Gordon Hamiltonian. If the displacement fluctuations were exactly Gaussian one would have
\begin{eqnarray}
\mathcal{A}(q)\Big|_{{\rm Gaussian}} & = A(\tau) q^2
\end{eqnarray}
for all $q$. Interestingly, the present more detailed calculation confirms
that property for $q<1$. For   $q=1$,
the amplitude $\mathcal{A}(q)$ of the correlation function $C(q,r)$  changes abruptly to
\begin{eqnarray}\label{conjecture}
\mathcal{A}(q=1)=6 \tau^2\ ,
\end{eqnarray}
instead of $A(q=1^-)=2\tau^2$. Hence our results both for $\eta(q),$ and for  ${\cal A}(q=1),$
  show deviations of the probability distribution of
 $\theta$ from a Gaussian. Since the functions
 $\eta(q)$ and ${\cal A}(q)$ are both increasing functions of $q$, the correlation
 function $C(q,r)$ is a decreasing function of $q$ for $0 \leq q \leq 1$. The amplitude ${\cal A}(q)$ jumps from $2 \tau^2$ at $q=1^-$ to
 $6 \tau^2$ at $q=1$ resulting in a much faster decay. As a precursor of this effect, $\eta(q)$  diverges as $q$ approaches unity. Such a
resonance should be visible in Bragg scattering experiments, once
the scattering wavevector orthogonal to the surface matches $G_0=2\pi/a$.

For $q>1$ and $T<T_{c}$, one naively expects, both for $\eta(q)$ and ${\cal A}(q)$, additional resonances at wave vectors that
are integer multiples of $G_0$, and that the screening mechanism which operates for $T>T_{c}$ in Eq.~(\ref{toner}) is also important there.
While a preliminary study indicates that this is the case, the detailed study of $q>1$ is more complicated and
deferred to a future publication. \cite{led+unpub}

It is interesting to note that the case of integer $q$ is relevant for the
system of two-dimensional free fermions in a disordered
potential. \cite{Guruswamy+00,LeDoussal+07} This model
and the present random-phase sine-Gordon model are in correspondence via bosonization. More precisely,
$C(q=1,r)$ can be obtained as a four-fermion correlation function.
We have performed that calculation, and found the result  to  agree with
our  expression (\ref{conjecture}) for $q=1$. A more general study of $4q$ fermion correlator, that would allow us to study higher integer values of $q$ in the correlation function, is in progress. The calculation and results will be presented elsewhere. \cite{led+unpub}

This article is organized as follows. In Sec.~\ref{s:model} we define the model. In Sec.~\ref{s:gen+con-cor} we introduce a formalism that enables us to produce a controlled  expansion for the correlation function of interest. In Sec.~\ref{s:eval-cor-func} we evaluate this correlation function.  This is followed, in Sec.~\ref{s:result+conclusion}, by a discussion of the consequences,  and conclusions. The detailed calculation of several involved integrals and some other technical details are presented in the appendixes.

\section{Model and the phase diagram}\label{s:model}

We consider the two-dimensional XY model for a real displacement field $\theta(x) \in (-\infty,\infty)$, without vortices, and in the presence of a random symmetry-breaking field. In the realization of the model as a fluctuating surface of the crystal, $\theta$ describes the two-dimensional height of the surface. We have already considered the same model in an earlier publication \cite{Ristivojevic+12}; here we repeat the  necessary definitions for the present study. The model is defined via its Hamiltonian
\begin{align}\label{H}
H=\int\dif^2 x\left[\frac{\kappa}{2}\left(\nabla_x\theta\right)^2-h\cdot\nabla_x\theta- \frac{1}{a}\left(\xi e^{i\theta}+\mathrm{h.c.}\right)\right],
\end{align}
where $\kappa$ is the elastic constant, $a$ the lattice constant that provides a short-length-scale cutoff, and $h(x)$ and $\xi(x)$ are quenched Gaussian random fields, the first one real and the other complex. Their nonzero correlations are given by
\begin{align}\label{hihj}
&\overline{h^i(x)h^j(y)}=  T^2 \frac{\sigma}{2 \pi} \delta^{ij}\delta(x-y),\\
&\overline{\xi(x)\xi^*(y)}=T^2 \frac{g}{2 \pi}\delta(x-y),
\end{align}
where $i,j\in\{1,2\}$ denote the components of $h$ and $T$ is the temperature.\footnote{A real crystal has a finite correlation length $\xi_B$ for the bulk translational order. Here we assume delta correlated disorder and therefore capture physics at length scales larger than $\xi_B$.} Note that the disorder $h(x)$ must be introduced as it is generated by the symmetry-breaking field under coarse graining. We denote disorder averages by an overline. Depending on the context, $x$ and $y$ will be used either to denote two-dimensional coordinates (as in the previous equations) or as their norms, i.e., $x$ stands either for $\mathbf{x}$ or $|\mathbf{x}|$. We emphasize here that the argument of the exponent in Eq.~(\ref{H}) should in principle be $e^{i G_0\theta}$ with $G_0$ being the smallest nonzero reciprocal lattice vector of the crystal normal to its surface under consideration. For simplicity we have set it to unity, thus measuring the displacement field in units of $1/G_0$.

We use the replica method to treat the disorder. \cite{Giamarchi} Introducing the replicated fields $\theta_\alpha, \alpha=1\ldots n$, where by greek indices $\alpha,\beta,\ldots,$ we denote replica indices, the replicated Hamiltonian reads
\begin{align}\label{Hrep}
H^{\rm rep}=H_0^{\rm rep}+H_1^{\rm rep},
\end{align}
with the harmonic part
\begin{align}\label{H0}
\frac{H_0^{\rm rep}}{T}=&\sum_{\alpha\beta}\int\dif^2x\Big\{ \frac{\kappa}{2T}\delta_{\alpha\beta}\left[ (\nabla_x\theta_\alpha)^2+m^2(\theta_\alpha)^2\right]\notag\\
&-\frac{\sigma}{4 \pi}\nabla_x\theta_\alpha \cdot\nabla_x\theta_\beta\Big\}.
\end{align}
The mass $m$ is introduced  as an infrared cutoff.
We perform calculations with finite $m$ and study the limit $m\to 0$ at the end. The system size is infinite throughout the paper. The anharmonic part reads
\begin{align}
\label{H1}
\frac{H_1^{\rm rep}}{T}=-\frac{g}{2 \pi a^2}\sum_{\alpha\beta}\int\dif^2x\cos(\theta_\alpha-\theta_\beta).
\end{align}

We start by computing the correlation function for the harmonic part (\ref{H0}),  i.e., for $g=0$. One obtains \cite{Ristivojevic+12}
\begin{align}\label{cfcf}
G_{\alpha\beta}(x)=\langle \theta_\alpha(x)\theta_\beta(0)\rangle=\delta_{\alpha\beta}G(x)+G_0(x),
\end{align}
where $\langle\ldots\rangle$ denotes an average over thermal fluctuations, while at small distances $|x|\ll m^{-1}$ we obtain:
\begin{gather}\label{G-smallx}
G(x)=-(1-\tau)\ln \left[c^2 m^2(x^2+a^2)\right ],\\
\label{G0x}
G_0(x)=-2\sigma(1-\tau)^2\ln\left[e c^2m^2(x^2+a^2)\right]. \end{gather}
Here  we have introduced the ultraviolet regularization by the parameter $a$ and $c=e^{\gamma_{{E}}}/2$ with $\gamma_{{E}}$ being the Euler constant. The dimensionless parameter
\begin{align}
\tau=1-T/T_{c}
\end{align}
measures the distance from the critical super-roughening temperature
\be T_{c}=4\pi\kappa.\ee

The model studied here possesses an important symmetry, the statistical tilt symmetry (STS), i.e., the non-linear part $H_1^{\rm rep}$ is invariant under the change $\theta_\alpha(x) \to \theta_\alpha(x) + \phi(x)$ for an arbitrary function $\phi(x)$. As discussed in many works,
\cite{Schulz+88,Hwa+94,Carpentier+97,LeDoussal2010} this implies that $G_0(x)$ does not appear to any order in perturbation theory in $g$ in the calculation of, e.g., the effective action. \cite{Ristivojevic+12}

\begin{figure}
\includegraphics[width=0.7\columnwidth]{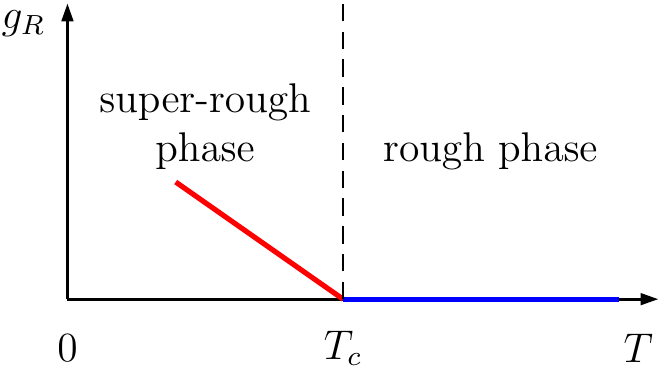}
\caption{Phase diagram for the model (\ref{H}). The high-temperature rough phase is characterized by zero fixed point values of the anharmonic $g_{R}$ term, while the low-temperature super-rough phase has a line of fixed points for $g_{R}$ having finite values. We are primarily interested to evaluate $C(q,r)$ in the low-temperature phase, for $T<T_{c}$.}\label{figdiagram}
\end{figure}

Let us summarize the  one-loop renormalization group equations for the model (\ref{H}).\cite{Cardy+82,Hwa+94,Carpentier+97,Ristivojevic+12} In terms of the scale
$\ell:=-\ln m$ they read
\begin{gather}
\label{RGtau}
\frac{\dif\tau}{\dif\ell}=0, \\
\label{RGgfinal}
\frac{\dif g_{R}}{\dif\ell}=2\tau g_{R} - 2 g_{R}^2,\\
\label{RGsigmafinal}
\frac{\dif\sigma_{R}}{\dif\ell}=\frac{1}{2} g_{R}^2,
\end{gather}
where the subscript $_{R}$ denotes the renormalized parameters that flow with the scale. Equation (\ref{RGtau}) is an exact result at all orders due to the above mentioned STS. We also see that the parameter $\sigma_{R}$ does not enter any equation (again due to STS), apart from being created by $g_{R}$ as one goes to larger scales. From Eq.~(\ref{RGgfinal}) one  reads off that the critical temperature is at $\tau=0$. For $T>T_{c}$ we have scaling of $g_{R}$ to zero at large scales which denotes the rough phase. At low temperatures $T<T_{c}$ one finds a line of nonzero fixed points for $g_{R}$ that determines the super-rough phase. We  see from the above scaling equations that $\sigma_{R}$  grows unboundedly  at all scales due to the non-zero fixed-point value for $g_{R}$;  hence $\sigma_{R}$ increases the  two-point correlation function from a rough logarithmic behavior in the high-temperature phase  to a super-rough square-logarithmic form for low temperatures.\cite{Toner+90} In Fig.~\ref{figdiagram} we show the phase diagram for the model (\ref{H}). The equations at two-loop order have been derived and studied in Ref.~\onlinecite{Ristivojevic+12}. They do not quantitatively change the conclusions from the one-loop analysis, but give further insight into the two-point correlation function in the low-temperature phase.

\section{Generator of connected correlations}
\label{s:gen+con-cor}

The main goal of the present study is to compute the exponential correlation function
\begin{align} \label{G1}
C(q,r) =& \overline{\langle e^{iq\left[\theta(r)-\theta(0)\right]}\rangle}_H\notag\\
=&\left\langle e^{iq\left[\theta_\gamma(r)-\theta_\gamma(0)\right]}\right\rangle_{H^{\mathrm{rep}}}
\end{align}
for the model (\ref{H}) in the low-temperature phase. For actual calculations we use the second line of Eq.~(\ref{G1}) that is obtained using replicas. By $\gamma$ we denote a particular replica index.

In order to calculate the correlation function (\ref{G1}) it is useful to calculate the generator of connected correlations ${W}(J)$. It is defined as \cite{Zinn-Justin}
\begin{align}\label{W}
e^{{W}(J)}=\int\Dif[\varphi]\, e^{-H(\varphi)/T+J\varphi},
\end{align}
where $H(\varphi)/T=H_0(\varphi)/T+gV(\varphi)$ is the reduced Hamiltonian. Here $H_0$ is the quadratic part and $V$ some perturbation. After performing the shift of the field as $\varphi=\chi+GJ$, one obtains $H_0(\varphi)-\varphi J=H_0(\chi)-H_0(GJ)$, where $G$ is the propagator defined by $H_0/T=\frac{1}{2}\langle\varphi|G^{-1}|\varphi\rangle$ where we use the compact notation. Equation (\ref{W}) then becomes
\begin{align}\label{Gamma}
e^{W(J)}=&\int\Dif\chi \exp\left[-\frac{H_0(\chi)}{T}+\frac{H_0(GJ)}{T}\notag-g V(GJ+\chi)\right]\notag\\
=&Z_0\exp\left[\frac{H_0(GJ)}{T}\right] \left\langle\exp\left[-gV(GJ+\chi)\right]\right\rangle^{\chi}.
\end{align}
Using the cumulant expansion we obtain a perturbative expansion for $W(J)$ that reads
\begin{align}\label{Wcumulant}
W(J)=&\ln Z_0+\frac{H_0(GJ)}{T} -g\left\langle V(GJ+\chi)\right\rangle^{\chi}\notag\\
&+\frac{g^2}{2}\left\langle V^2(GJ+\chi)\right\rangle^{\chi}_c+\mathcal{O}(g^3).
\end{align}

In the following we calculate a perturbative expansion for the functional $W(J)$ using the previous expression (\ref{Wcumulant}). Having evaluated $W(J)$, one can immediately obtain all  connected correlation functions by differentiating the left- and right-hand side of its defining equation [equivalent to Eq.~(\ref{W})]
\begin{align}\label{Wconnection}
W(J)=\ln Z+\ln\left\langle \exp(J\varphi)\right\rangle_H
\end{align}
an arbitrary number of times with respect to the source field $J$ and setting $J$ to zero at the end. For potentials that are even, i.e., when $V(\varphi)=V(-\varphi)$, only  terms with an even number of $\varphi$ fields have non-zero correlations.\\

\subsection{Expressions for $W(J)$}

Using the derived formula (\ref{Wcumulant}) and applying it to our replicated Hamiltonian (\ref{Hrep}) we obtain
\begin{align}\label{W(J)pert}
W(J)=\ln Z_0+ \frac{H_0^{\mathrm{rep}}(GJ)}{T}+W_1+W_2+\mathcal{O}(g^3),
\end{align}
where the term $W_j$ comes from the corresponding term of Eq.~(\ref{Wcumulant}) proportional to $g^j$. Either directly calculating or using the obtained results for the effective action from Ref.~\onlinecite{Ristivojevic+12} and the correspondence from Appendix~\ref{appendix:GammaWconnection} we obtain the final results for $W_1$ and $W_2$. We note that in models that have STS it was shown that the two-replica part of the functional $W$ and of the effective action $\Gamma$ are identical up to the replacement $\theta\to GJ$, \cite{LeDoussal2006b,LeDoussal2010} see Appendix \ref{appendix:GammaWconnection}.

The lowest order term reads
\begin{align}
\label{W1}
W_1=&\frac{g}{2 \pi a^2} e^{-G(0)}\notag\\ &\times\sum_{\alpha\beta}\int \dif^2x \cos\left\{\int\dif^2 y G(x-y)\left[J_{\alpha}(y)-J_\beta(y)\right]\right\}\notag\\
=&0.
\end{align}
The last line is obtained in the limit we need, that is $m\to 0$, when the power law $e^{-G(0)}=(c^2m^2a^2)^{1-\tau}$ compensates the logarithmic divergence of $G(x)$ under the cosine. The second order term $W_2$ contains several different contributions. In Appendix~\ref{appendix:W2} we give a complete expression. Here we use only the final result taken in the limit $m\to 0$. It reads

\begin{widetext}
\begin{align}\label{W2final}
W_2=&\frac{g^2}{8\pi^2 a^4}\sum_{\alpha\beta}\int\dif^2x\dif^2y \left[\frac{a^2}{(x-y)^2+a^2}\right]^{2(1-\tau)}\cos\left\{\int\dif^2 z\left[(1-\tau)\ln\frac{(y-z)^2+a^2}{(x-z)^2+a^2}\right] \left[J_\alpha(z)-J_\beta(z)\right]\right\}.
\end{align}
\end{widetext}

\subsection{Special source field}

Using a source field of the form
\begin{align}\label{Jspecial}
J_\alpha(x)=i q \left[\delta(x-r)-\delta(x)\right]\delta_{\alpha\gamma},
\end{align}
in Eq.~(\ref{Wconnection}), we obtain a relation between the correlation function (\ref{G1}) and the functional $W(J_\alpha)$. It reads
\begin{align}\label{mainconnection}
\left\langle e^{i q \left[\theta_{\gamma}(r)-\theta_{\gamma}(0)\right]}\right\rangle_{H^{\mathrm{rep}}}=\frac{1}{Z} e^{W(J_\alpha)}.
\end{align}
The left-hand side of Eq.~(\ref{mainconnection}) is  the correlation function $C(q,r)$, while the right-hand side of it should be evaluated. Here we perform a perturbative calculation in the anharmonic coupling [defined in Eq.~(\ref{H1})] of $W(J_\alpha)$, using the expansion (\ref{W(J)pert}).

\begin{figure}
\includegraphics[width=0.6\columnwidth]{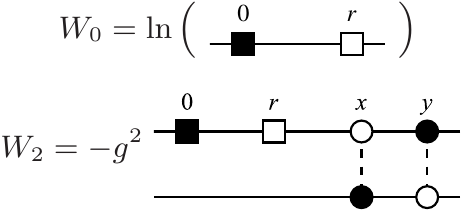}
\caption{A graphical representation of the non-vanishing contributions to the perturbative expansion. The white and black squares stand for the observable $e^{iq \theta_{\gamma}(r)} $ (with charge $+q$) and $e^{-iq \theta_{\gamma}(0)} $ (with charge $-q$) respectively. The replica indices are indicated by a solid line. The interaction  $\sum_{\alpha\neq \beta}e^{i [\theta_{\alpha}(x)-\theta_{\beta}(x)]} $ is coded with a white and black circle, joined by a dashed line. There has to be as many black as white points and circles in each given replica (charge neutrality). Charges $q_{1}$ at $x_{1}$ and $q_{2}$ at $x_{2}$ interact with a multiplicative factor (Boltzmann weight) of $|x_{1}-x_{2}|^{2q_{1}q_{2}T/T_{c}}$; they   ``attract'' for opposite signs, leading to a divergent contribution as their distance goes to zero. This allows for an intuitive visualization of the corresponding divergences in the perturbation expansion. }
\label{f:graphs}
\end{figure}

For the source field (\ref{Jspecial}), the lowest-order term in the expansion of $W(J_\alpha)$  becomes
\begin{align}\label{W0}
W_0&=\frac{H_0^{\mathrm{rep}}(GJ)}{T}=\sum_{\alpha\beta}\int\dif^2 x\dif^2 y J_\alpha(x)\frac{G_{\alpha\beta}(x-y)}{2}J_\beta(y)\notag\\
&= -q^2(1-\tau)\left[1+2(1-\tau)\sigma\right] \ln\left(\frac{r^2+a^2}{a^2}\right).
\end{align}
A graphical interpretation is given in Fig.~\ref{f:graphs}.

The second nonzero term of $W(J_\alpha)$ is obtained by employing the source field (\ref{Jspecial}) in Eq.~(\ref{W2final}), and by making use of the summation formula over replica indices  $\sum_{\alpha,\beta=1}^n\cos[A(\delta_{\alpha\gamma}-\delta_{\beta\gamma})] =n^2+2(1-n)(1-\cos A)$. In the  limit $n\to 0$, we obtain
\begin{align}\label{W2int}
W_2&=-g^2\left(\frac{r}{a}\right)^{4\tau}I(r,a,q,\tau),
\end{align}
where we have defined the following integral:
\begin{align}\label{Integral}
I(r,a,q,\tau):=&\frac{r^{-4\tau}}{4\pi^2}\int\dif^2 x\dif^2 y\frac{1}{\left[a^2+(\mathbf{x}-\mathbf{y})^2\right]^{2(1-\tau)}}\notag\\ &\times\left\{\left[\frac{(\mathbf{x}-\mathbf{r})^2+a^2}{(\mathbf{y}-\mathbf{r})^2+a^2}\  \frac{\mathbf{y}^2+a^2}{\mathbf{x}^2+a^2}\right]^{q(1-\tau)}-1\right\}.
\end{align}
It is an even function of $q$, and as a function of $r$ and $a$ depends only on $r/a$, provided it is convergent.

Finally, using Eqs.~(\ref{W(J)pert}), (\ref{W0}), (\ref{W2int}), and (\ref{mainconnection}), as well as
the result $Z_0/Z=1$ valid in the replica limit $n\to 0$, we obtain the final (yet unevaluated) result for the correlation function (\ref{G1}),
\begin{align}\label{G1final}
C(q,r)=& \exp\left(-q^2(1-\tau)[1+2(1-\tau)\sigma]\ln\left(\frac{r^2+a^2}{a^2}\right) \right)\notag\\ &\times\exp\left(-g^2\left(\frac{r}{a}\right)^{4\tau} I(r,a,q,\tau)\right)\exp\left({\mathcal{O}(g^3)}\right).
\end{align}

Equations (\ref{Integral})-(\ref{G1final}) are the starting point of the evaluation of the correlation function (\ref{G1}). One should
read them as the result of perturbation theory to leading order in the bare coupling $g$. However, if we  re-express
$g$ in terms of the renormalized one $g_{R}$, we obtain the result to leading order in $\tau$, the distance from the transition.
They are connected by the relation\cite{Ristivojevic+12} $g=g_{R}[1+\mathcal{O}(\tau)]$. In the massless limit, equivalently, at large distances, $r\gg a$, from the renormalization group equation (\ref{RGgfinal}) the renormalized coupling reaches its fixed point value $\tau+\mathcal{O}(\tau^2)$,
hence we can replace
\begin{eqnarray} \label{gtotau}
g \to \tau+\mathcal{O}(\tau^2)
\end{eqnarray}
in the evaluation of the large distance behavior. The behavior of $g_{R}$ at intermediate scales before reaching the fixed point
causes some non-universal behavior. It is easy to see, however, that it is confined to subdominant terms, such as the part of $\ln (r/a)$ that is proportional to $q^2$. As a consequence of flowing of $g_R$ toward the fixed point, the off-diagonal disorder becomes changed at large scales. Therefore by $\sigma'$ we denote the non-universal parameter that characterizes the off-diagonal disorder. We note that, to lowest order in $g$, we have $\sigma'=\sigma$. In order to emphasize the difference between the bare parameter $\sigma$, defined in Eq.~(\ref{hihj}), and the effective one at large scales $\sigma'$, we keep $\sigma'$ in final formulas, e.g., in Eq.~(\ref{ourc}).

Thus, to leading order, we only need the integral $I(r,a,q,\tau=0)$, that is a function of the ratio $r/a$ and $q$ only. It is clear from the definition (\ref{Integral}) that for $\tau=0$ this integral is  convergent for any $q$:
It is ultraviolet convergent thanks to the cutoff $a$ and infrared convergent thanks to the substraction
of $1$. There are however ultraviolet divergences when $a/r \to 0$,
which lead to a logarithmic behavior, and which will be analyzed in the next section.
Here we want to point out that while for $|q|<1$ they come
from the first power law factor, there is an additional ultraviolet divergence arising from
the second factor (containing $r$) when $q=1$. Hence we will mainly restrict to
$q \leq 1$, and discuss separately the cases $|q|<1$ and $q=1$.
It makes physical sense that at $q=1$ (and in general any integer $q$) the correlation function changes non-smoothly, as was already the case in high temperature phase [see the results (\ref{toner}) of \citet{Toner+90}].

\section{Evaluation of the correlation function}
\label{s:eval-cor-func}

In this section we evaluate the correlation function $C(q,r)$ defined by Eq.~(\ref{G1}), using its form (\ref{G1final}), for general values of the parameter $q$ in the region $0<q\le 1$. To achieve this, our task is to calculate the integral of Eq.~(\ref{Integral}). Exact evaluation of that integral is rather difficult. However, we only need the most dominant term for large $r/a$, when the expression of Eq.~(\ref{Integral}) is expected to take a universal form.

We calculate the integral using two different methods. The first one, which we term the  finite-$a$ method, works directly at $\tau=0$, but with a non-zero ultraviolet cutoff $a>0$. However
we are not able to obtain all the results using this finite-$a$ method. Hence we also use dimensional regularization, which we term the dimensional method; it is quite powerful,
although a little delicate in its interpretation. In that method one works directly at $a=0$,  but
keeps $\tau>0$, which renders the integral convergent. From the poles in $1/\tau$ one extracts the desired information.
A careful comparison between the obtained results is performed at the end.

\subsection{Finite-$a$ method}

We start the evaluation by setting $\tau=0$ and keeping $a$  finite in Eq.~(\ref{Integral}). This is  justified as we have $I(r,a,q,\tau)=I(r,a,q,0)+\mathcal{O}(\tau)$, so to lowest order in the distance from the transition we only need $I(r,a,q,0)$, as discussed above.

For small $q$, we expand it into a Taylor series as
\begin{align}\label{expansion}
I(r,a,q,0)=&\sum_{j=1}^\infty q^{2j} I_{j}(r/a),
\end{align}
where we have defined the family of integrals
\begin{align}\label{I1}
I_{j}(r)=&\frac{1}{4\pi^2(2j)!}\int\dif^2 x\dif^2 y\frac{1}{\left[1+(\mathbf{x}-\mathbf{y})^2\right]^{2}}\notag\\ &\times \ln^{2j}\left[\frac{(\mathbf{x}-\mathbf{r})^2+1}{(\mathbf{y}-\mathbf{r})^2+1}\ \frac{\mathbf{y}^2+1}{\mathbf{x}^2+1}\right].
\end{align}
The expansion (\ref{expansion}) contains only even powers of $q$, a consequence of the parity $I(r,a,q,0)=I(r,a,-q,0)$. A rather involved evaluation of the lowest-order term $I_1(r)$ of Eq.~(\ref{expansion}) is presented in Appendix~\ref{appendix:finite a integrals}. At small $q$, the result reads
\begin{align}\label{Ispec}
I(r,a,q,0)=q^2\ln^2\left(\frac{r}{a}\right)+q^2 \ln\left(\frac{r}{a}\right)+r(q),
\end{align}
where the even function $r(q)$ at small $q$ starts with a term proportional to $q^4$ and
contains all the desired higher-order terms proportional to $q^{2 j}$, $j \ge 2$.

\begin{figure}
\includegraphics[width=0.9\columnwidth]{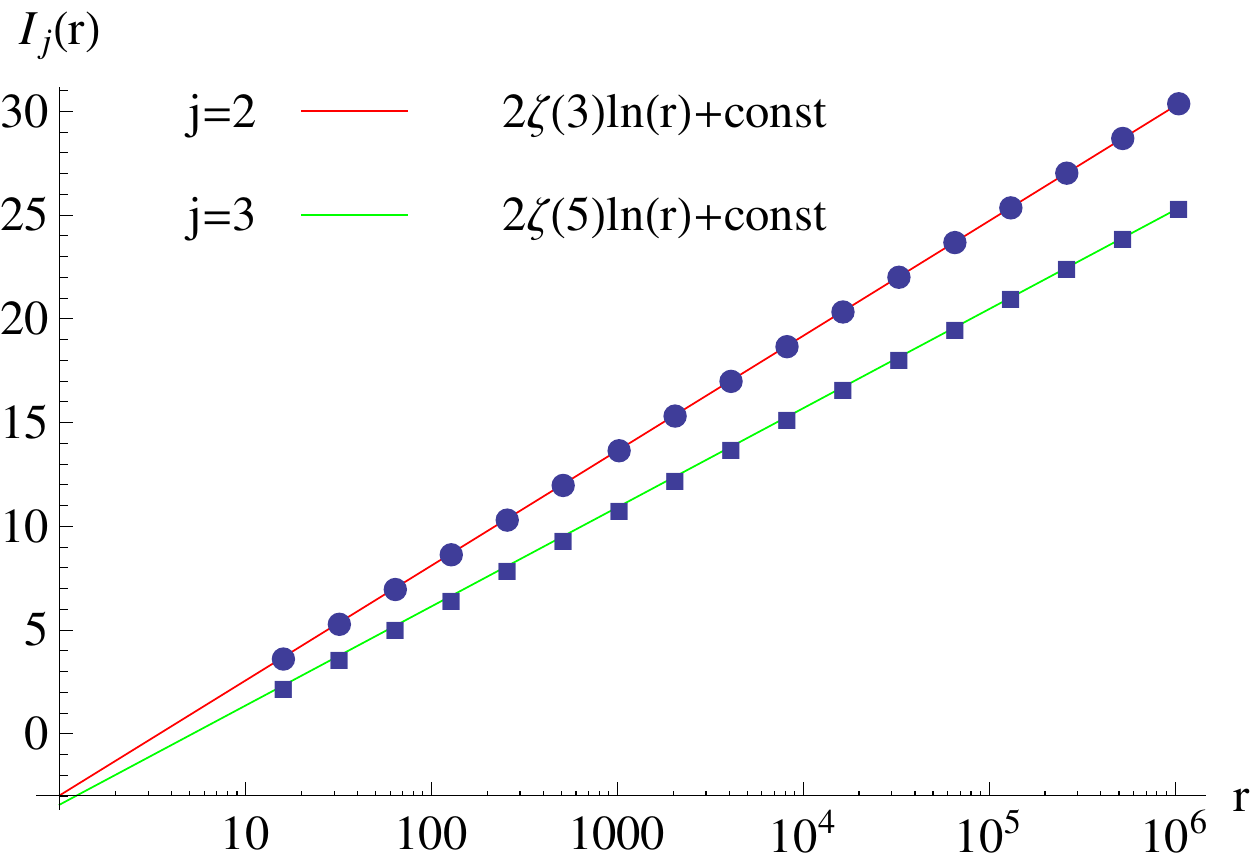}
\caption{Numerical evaluation of Eq.~(\ref{I1}) for $j=2,3$. At large $r$, we observe that the data belong to a straight line on a log-linear plot. The numerical coefficients [$2\zeta(3)$ and $2\zeta(5)$ for $j=2$ and $j=3$, respectively] in front of the logarithms in the fitted straight lines are analytically obtained by the dimensional method, see the main text.}\label{figI1}
\end{figure}

To determinate $r(q)$ defined in Eq.~(\ref{Ispec}) we begin by a numerical evaluation. In Fig.~\ref{figI1} we show the results for $I_j(r)$ for $j=2,3$ that determine the two lowest-order terms of $r(q)$ at small $q$, see Eqs.~(\ref{expansion}) and (\ref{Ispec}). We notice that the data points for $I_2(r)$ and $I_3(r)$ appear to be on a straight line on a log-linear plot, which shows that these two functions are well described by a logarithm $\ln(r)$ with $j$-dependent prefactors [$2\zeta(3)$ and $2\zeta(5)$ in the present case]. This leaves us with a hint that all $I_j(r)$ for $j>1$ might asymptotically have  a logarithmic behavior in the limit $r\gg 1$. That would determine the anomalous exponent for the $C(q,r)$ correlation function.

We gain further knowledge about $I(r,a,q,0)$ by considering the special case $q=1$. The calculation presented in Appendix~\ref{appendix:finite a integrals} reveals the result
\begin{align} \label{res0}
I(r,a,1,0)=3\ln^2\left(\frac{r}{a}\right)-2\ln\left(\frac{r}{a}\right) +\mathcal{O}\left(\frac{a^2}{r^2}\right),
\end{align}
showing that the coefficient $q^2$ of the leading squared logarithmic term from the small-$q$ expansion of Eq.~(\ref{Ispec})  changes at $q=1$.
As discussed above this is expected from physical considerations. To confirm this result for $q=1$
we have performed a calculation in the fermionic version of the model that allows us to treat integer $q$ only. It  confirms our result for $q=1$. The calculation and results will be presented elsewhere.\cite{led+unpub}
Note that at $q=1$ the prefactor in front of $\ln(r/a)$ term in Eq.~(\ref{res0}) takes this value for the particular cutoff procedure we use, but is not expected to be universal in general.

Motivated by the above analysis, we introduce the anomalous exponent $\eta(q)$ and the amplitude $\mathcal{A}(q)$ and assume a general form for the correlation function at  $r\gg a$ given by Eq.~(\ref{G1form}), where
\begin{gather}\label{eta-def}
\eta(q)=2q^2(1-\tau)[1+2(1-\tau)\sigma]+g^2\eta_g(q)+\mathcal{O}(g^3),\\
\label{A-def}
\mathcal{A}(q)=\alpha(q) g^2+\mathcal{O}(g^3).
\end{gather}
For the unknown coefficients in the previous two expressions, so far we have established the following results:
\begin{gather}
\eta_g(1)=-2,\\
\alpha(q)=\begin{cases}
2q^2,& q\ll 1\\
6, &q=1
\end{cases}.
\end{gather}

\subsection{Dimensional method}

To evaluate Eq.~(\ref{Integral}) using the dimensional method, we take  the limit $a\to 0$ of the integrand, i.e., we consider
\begin{align}\label{Idim}
I(r,a,q,\tau)=&I(r,0,q,\tau)+\mathcal{O}(a^2)\notag\\
=&I(1,0,q,\tau)+\mathcal{O}(a^2),
\end{align}
where the second line is a trivial consequence of rescaling the coordinates by $r$, provided the
integral is convergent. For $\tau >0$ the integral $I(1,0,q,\tau)$ is ultraviolet convergent as long as $|q|(1- \tau)<1$; it
remains infrared convergent if $\tau$ is not too large ($\tau<1/2$). We thus first discuss $|q|<1$. From the poles of the evaluated expression in the limit $\tau\to 0$
we will infer the behavior of the integral at nonzero $a$, as is shown below. We rewrite
\begin{equation}
I(1,0,q,\tau)=\frac{1}{4\pi^2}J(q(1-\tau),\tau),
\end{equation}
where we defined
\begin{equation}\label{40}
\\ J(p,\tau) := {\rm FP} \int\dif^2 x\,\dif^2 y\, |\mathbf{x}-\mathbf{y}|^{4(\tau-1)}\left[\frac{|\mathbf{x}-\mathbf{e}|
|\mathbf{y}|}{|\mathbf{y}-\mathbf{e}||\mathbf{x}|}\right]^{2p}.
\end{equation}
Here \(\bf e\) is an arbitrary unit vector, and
${\rm FP}$ means ``finite part" in the sense of dimensional regularization
[in some domain of $q$ this finite part is achieved by the subtraction $1$ in Eq.~(\ref{Integral}), however
it can be given a more general meaning in terms of analytical continuation in the parameters $q,\tau$].
The evaluation of the ensuing \(2\times2\)-dimensional integral is complicated
and the details are presented in Appendix \ref{appendix:integral}.
Let us recall here the main idea, which goes back to Dotsenko and Fateev \cite{DotsenkoFateev1984,DotsenkoFateev1985}: The integral (\ref{40}) can be thought of as an integral over the complex plane, both for  $x$ and $y$. The integral over, say $x$ over the complex plane can be decomposed into two real contour integrals, over $x$ and its complex-conjugate $\bar x$. Noting that (\ref{40}) can be written in the form \begin{align}
& \int\dif^2 x\,\dif^2 y\, (1-x)^{ p} y^{ p} x^{2 \tau -1} (1-y)^{2 (\tau
   -1)} (1-x y)^{- p} \nn\\
   &\qquad \ \ \ \ \, \times (1-\bar x)^{ p} \bar y^{ p} \bar x^{2 \tau -1} (1-y)^{2 (\tau
   -1)} (1-\bar x \bar y)^{- p}\ , \label{41}
\end{align}
it is suggestive that deforming the contour integrals over \(x\) and \(\bar x\) to lie on the real axis, the resulting integral will take the form \begin{displaymath}
J(p,\tau) = \sum_i B_i J_i J_i' \ .
\end{displaymath}
The $B_i$'s are phase-factors -- typically deforming a contour around a branch cut gives a sine-function of $2\pi $ times the power at the branch cut. The $J_i$  are second-generation hypergeometric functions ${}_3F_2$, e.g., the first line of Eq.~(\ref{41}) will naturally lead to
$$J_i = \int_0^1 \dif x \int_0^1 \dif y\, (1-x)^{ p} y^{ p} x^{2 \tau -1} (1-y)^{2 (\tau
   -1)} (1-x y)^{- p}\ .
$$
Note  that the integral above is restricted, both for $x$ and $y,$ to the interval \([0,1]\); but the domains \([-\infty,0]\) and \([1,\infty]\) also contribute; the latter can then be transformed back to the interval $[0,1]$, leading to integrals of the same form, but with different coefficients. This form can be recognized in the  exact result given in Appendix \ref{appendix:integral}, formula (\ref{Jguida}).

The final result for  $0<q<1$ and  small $\tau$ reads
\begin{align}\label{I(1)}
I(1,0,q,\tau)=&\frac{q^2}{8\tau^2}+\frac{q^2\left[1-2\gamma_{E} -\psi(q)-\psi(-q)\right]}{4\tau}\notag\\
&+\mathcal{O}(\tau^0),
\end{align}
where $\psi(x)=\Gamma'(x)/\Gamma(x)$ is the digamma function. The coefficients of the poles in Eq.~(\ref{I(1)}) contain all information needed for the determination of the anomalous exponent (\ref{eta-def}) and the amplitude (\ref{A-def}). The precise method to extract the information from these poles is given in Appendix \ref{app:poles}. Here we give a more intuitive derivation.

Starting from Eq.~(\ref{W2int}), which is contained in the correlation function (\ref{G1final}), we expand $(r/a)^{4\tau}I(r,a,q,\tau)$ at small $\tau$. Using Eqs.~(\ref{Idim}) and (\ref{I(1)}), for $0<q<1$ we find the result
\begin{align}\label{translation}
\left(\frac{r}{a}\right)^{4\tau}I(r,a,q,\tau)=q^2\ln^2\left(\frac{r}{a}\right) +\eta_g(q)\ln\left(\frac{r}{a}\right) +\mathcal{O}(\tau),
\end{align}
where the prefactor of the logarithmic term, i.e., the nontrivial part of the anomalous exponent, reads
\begin{align}\label{etagf}
\eta_g(q)=&q^2[1-2\gamma_{E}-\psi(q)-\psi(-q)].
\end{align}
Equation (\ref{translation}) could also be understood as the final expression for $I(r,a,q,\tau=0)$ obtained via the dimensional method. At small $q$, one can notice the agreement between the two results (\ref{Ispec}) and (\ref{translation}) that are obtained using quite different methods.

The expression (\ref{translation}) should be taken with special attention. The translation of the dimensional-method result (\ref{I(1)}) by performing a naive expansion at small $\tau$ of the left hand side of Eq.~(\ref{translation}) would contain extra terms
\begin{align}\label{extra}
\frac{q^2}{8\tau^2}+\frac{1}{4\tau}\left[2q^2\ln\left(\frac{r}{a}\right)+\eta_g(q)\right],
\end{align}
on the right hand side. These two terms are divergent for $\tau\to 0$. The terms (\ref{extra}) formally appear only because we performed a small $a$ expansion in Eq.~(\ref{Idim}), i.e., we set $a=0$ in the integrand. The left-hand side of Eq.~(\ref{translation}) should be finite at $\tau=0$, and thus the right-hand one as well. Therefore, the extra terms (\ref{extra}) should be omitted when one relates the dimensional result (\ref{I(1)}) to the expression for $I(r,a,q,\tau=0)$. We again emphasize that  a more rigorous derivation is given in Appendix \ref{app:poles}, while Appendix~\ref{app:connection2} we give an illustrative example.

\section{Final results and conclusion}
\label{s:result+conclusion}

Using Eqs.~(\ref{G1final}) and (\ref{translation}) we obtain the final result for the correlation function
\begin{align}\label{Cdef}
C(q,r) =& \overline{\langle e^{iq\left[\theta(r)-\theta(0)\right]}\rangle}_H,
\end{align}
at $q\le 1$, for the model (\ref{H}).
It takes the form
\begin{align}\label{Cfinalform}
C(q,r)\simeq\left(\frac{a}{r}\right)^{\eta(q)}
\exp\left[-\frac{1}{2}\mathcal{A}(q)\ln^2(r/a)\right],
\end{align}
where the amplitude is given by
\begin{align}\label{amplitudeFINAL}
\mathcal{A}(q)=\begin{cases}
2q^2 \tau^2+\mathcal{O}(\tau^3),& q<1\\
6\tau^2+\mathcal{O}(\tau^3), &q=1
\end{cases}.
\end{align}
Here and below we use the replacement $g \to \tau$ as discussed in Eq.~(\ref{gtotau}), i.e., the fixed point of the renormalization group.

The anomalous exponent of Eq.~(\ref{Cfinalform}) reads
\begin{gather}\label{etaFINAL}
\eta(q)=2q^2(1-\tau)[1+2(1-\tau)\sigma'] +\tau^2\eta_g(q)+\mathcal{O}(\tau^3),
\end{gather}
where its nontrivial part determined by the anharmonic coupling reads
\begin{gather}\label{etagfinal}
\eta_g(q)=\begin{cases}
q^2[1-2\gamma_{E}-\psi(q)-\psi(-q)], &q<1\\
-2,& q=1\end{cases}
\end{gather}
Here $\gamma_{E}$ is the Euler constant. This produces the result (\ref{etafinal}) displayed in the
Introduction where we have separated the universal part, $\widetilde \eta(q)$, of $\eta(q)$
that starts at order $q^4$, from the part proportional to $q^2$ that is non-universal.

The anomalous exponent $\eta(q)$ rapidly increases as $q$ increases from zero to one.
Thus, $C(q_1,r)>C(q_2,r)$ for $0<q_1<q_2<1$, meaning the correlation function decreases as $q$ increases.
When $q$ approaches unity, $\eta(q) \simeq 1/(1-q)$ becomes very large and makes the correlation function $C(q,r)$ decay much faster, see Eq.~(\ref{Cfinalform}).
This is a precursor  of the more drastic effect that happens at $q=1$ where the amplitude $\mathcal{A}(q)$ jumps
(from $2q^2 \tau^2$ at $q\to 1^-$ to $6\tau^2$ at $q=1$). These dips in the correlation
function $C(q,r)$ are a remarkable and unique feature of the super-rough phase. Note that near $q=1$
there is a growing length scale
\begin{align}
\xi_q\approx a \exp\left(\frac{1}{1-q}\right),
\end{align}
below which the $\eta(q) \ln(r/a)$ term in the exponential
in (\ref{Cfinalform}) is larger than the asymptotic $\ln^2(r/a)$ term.

Our  result (\ref{etagfinal}) can also be expanded at small $q$ as
\begin{align}
\eta_g(q)=q^2+2\zeta(3)q^4+2\zeta(5)q^6+\mathcal{O}(q^8).
\end{align}
This expansion precisely determines the prefactors for the family of integrals of Eq.~(\ref{I1}).
For $j\ge2$ we find $I_j(r)=2\zeta(2j-1)\ln(r)+\mathcal{O}(r^0)$, which  explains the data of Fig.~\ref{figI1}. This provides a confirmation that we have correctly extracted the amplitudes from the dimensional method (at least for small $q$).

Our result for the correlation function (\ref{Cdef}) enables us to calculate the leading large-distance behavior of all higher powers of the connected correlation functions in the super-rough phase, i.e., for $T<T_{c}$ (see Fig.~\ref{figdiagram}). Using the cumulant expansion formula $\ln\langle \exp[A]\rangle=\sum_{j=1}^\infty \langle A^j\rangle_c/j!$, after expanding (the logarithm of) Eqs.~(\ref{Cdef}) and (\ref{Cfinalform}) at small $q$, one obtains
\begin{gather}
\overline{\left\langle\left[\theta(r)-\theta(0)\right]^{2j-1}\right\rangle}_c=0,\label{cum1}\\
\frac{(-1)^{j}}{(2j)!}\overline{\left\langle\left[\theta(r)-\theta(0)\right]^{2j}\right\rangle}_c= -2\tau^2\zeta(2j-1)\ln\left(\frac{r}{a}\right)\notag\\
\qquad\qquad\qquad\qquad\quad\ +\mathcal{O}\left(\tau^2\frac{r^0}{a^0},\tau^3\right)
\end{gather}
for $j>1$. On the contrary, for $j=1$ one finds
\begin{align}\label{superrough}
\overline{\langle\left[\theta(r)-\theta(0)\right]^2\rangle}= 2\tau^2\ln^2\left(\frac{r}{a}\right)+\mathcal{O}\left(\ln\left(\frac{r}{a}\right)\right).
\end{align}
Expression (\ref{superrough}) is the well-known result \cite{Carpentier+97,Ristivojevic+12,perret+12PhysRevLett.109.157205} for the model (\ref{H}). This is yet another way of obtaining the result that produced some controversies in the past, as discussed in Ref.~\onlinecite{Carpentier+97}. Confirming a recently obtained correction\cite{Ristivojevic+12} to the prefactor of the squared logarithm in Eq.~(\ref{superrough}) using the present method would require explicit evaluation of the ${\mathcal{O}(g^3)}$ term in Eq.~(\ref{G1final}), which is a formidable task beyond the scope of the present study.

Our main results directly apply to some physical systems, in particular to surfaces of crystals with quenched bulk disorder \cite{Toner+90} or to a vortex lattice confined to a plane. \cite{Hwa+94} In particular, the structure factor $S(q,\mathbf{k})=\int\dif^2 r\, C(q,r)\exp(i \mathbf{k}\;\mathbf{r})$ in the super-rough phase at $k=0$ is analytic and finite.\cite{Toner+90} However, we predict that it has sudden dips once the wave-vector $q$ becomes an integer multiple of the reciprocal lattice vector normal to the surface $G_0$ (set to unity here) of the bulk crystal;  we expect it to have characteristic jumps in the amplitude {\em and} the anomalous exponent not only at $q=1$ but rather at any integer $q$.

To summarize, the main results of the present study are given by Eqs.~(\ref{Cdef})-(\ref{etagfinal}), or equivalently Eq.~(\ref{etafinal}). They describe the behavior of the exponential correlation function in the super-rough phase, defined by Eq.~(\ref{Cdef}). We found a characteristic universal jump of the amplitude of the squared logarithmic term, at $q=1$, which occurs in conjunction with the drop of the anomalous exponent. The value of the amplitude at $q=1$ is in agreement with the result in the equivalent fermionic disordered model and will be published elsewhere.\cite{led+unpub} It would be very interesting to perform a numerical simulation, along the lines of Ref.~\onlinecite{perret+12PhysRevLett.109.157205}, to test the results of the present work.

\acknowledgments

We thank Leon Balents, Andrei Fedorenko, and Leo Radzihovsky for valuable discussions. This work is supported by the ANR Grant No.~09-BLAN-0097-01/2. Z.R.~acknowledges the hospitality of MPI-PKS Dresden where  parts of the manuscript were written and financial support by Ecole Polytechnique.

\appendix

\section{$W_2$ term}\label{appendix:W2}

In this appendix we provide the complete expression for the second-order term $W_2$ of the generator of connected correlations, see Eq.~(\ref{W(J)pert}). It reads

\begin{align}\label{Gamma2}
W_2=&\frac{1}{2}\left(\frac{g}{2\pi a^2} \right)^2 e^{-2G(0)} \sum_{s=\pm}\sum_{\alpha\beta\gamma} \int\dif^2x\dif^2y\nn\\
&\times  \bigg\{\frac{1}{2}A_1(x-y)\delta_{\alpha\gamma} \cos\left[(GJ_{\alpha\beta})(x)\right]\notag\\
&\quad+\left[A_2^s(x-y)\delta_{\alpha\gamma}+A_4^s(x-y)\right]\nn\\
&\quad\times  \cos\left[(GJ_{\alpha\beta})(x)-s(GJ_{\gamma\beta})(y)\right] \bigg\},
\end{align}
where
\begin{gather}\label{A1}
A_1(x)=4(2-e^{-G(x)}-e^{G(x)}),\\
\label{A2}
A_2^s(x)=1+e^{2sG(x)}-2e^{sG(x)},\\
\label{A4}
A_4^s(x)=2(e^{sG(x)}-1).
\end{gather}
Here we introduced the abbreviation
\begin{align}\label{abbrev}
(GJ_{\alpha\beta})(x)\equiv\int\dif^2 y G(x-y)\left[J_{\alpha}(y)-J_\beta(y)\right].
\end{align}
In the limit of zero mass, $m\to 0$, the only term that survives in Eq.~(\ref{Gamma2}) comes from the $A_2^+$ term and reads
\begin{align}
&W_2=\frac{g^2}{8\pi^2 a^4}\sum_{\alpha\beta}\int\dif^2x\dif^2y e^{2G(x-y)-2G(0)}\notag\\
&\times\cos\left\{\int\dif^2 z\left[G(x-z)-G(y-z)\right]\left[J_\alpha(z)-J_\beta(z)\right]\right\}.
\end{align}
The difference of propagators in the previous equation is well behaved in the limit $m\to 0$. Using the propagator (\ref{G-smallx}), we eventually obtain the expression (\ref{W2final}) of the main text.

\section{Connection between the effective action $\Gamma$ and $W$}\label{appendix:GammaWconnection}

The final formula (\ref{Wcumulant}) resembles the expression (A17) of Ref.~\onlinecite{Ristivojevic+12} for the effective action, provided in the latter one omits all the terms that explicitly contain the propagator $G$ and one multiplies all the terms in Eq.~(A17) by a factor $-1$ apart from the harmonic part $S_0$ (corresponding to $H_0/T$ in the present work). We note that in general, in models that have STS, the two-replica parts of the functional $W$ and $\Gamma$ are identical up to the replacement of the form $\theta\to GJ$ and a global minus sign, see Eqs.~(89) and (90) of Ref.~\onlinecite{LeDoussal2010}.

For the model studied in the present work, we calculated the effective action $\Gamma$ in Ref.~\onlinecite{Ristivojevic+12}; the term $\Gamma_2$ proportional to $g^2$ is given by Eq.~(25) there. We notice that in Ref.~\onlinecite{Ristivojevic+12} we used restricted summation over replica indices (all replica indices different), while in the present work we do not impose such restriction. As a consequence, if one wants to use the theorem of similarity between the functionals $W$ and $\Gamma$,\cite{LeDoussal2006b,LeDoussal2010} one should transform the expression for $\Gamma$ to contain summations over unrestricted replica indices only, before applying the theorem. On the contrary, if one directly uses the expression for $\Gamma_2$, Eq.~(25) of Ref.~\onlinecite{Ristivojevic+12}, its two-replica part corresponds to the two replica part of Eq.~(\ref{Gamma2}), provided one omits the linear term $-pG(x)$ in $\Gamma_2$ and performs the replacement $\theta\to GJ$, or written more explicitly:
\begin{align}\label{changesWG}
&\theta_\alpha(x)\to\sum_\gamma\int\dif^2 y G_{\alpha\gamma}(x-y) J_{\gamma}(y),\\
\label{changesWG1}
&\theta_\alpha(x)-\theta_\beta(x)\to\int\dif^2 y G(x-y)\left[J_{\alpha}(y)-J_\beta(y)\right]\notag\\
&\quad\quad\quad\quad\quad\quad\equiv(GJ_{\alpha\beta})(x).
\end{align}
In the last expression we explicitly used the fact that $G_{\alpha\beta}(x)=\delta_{\alpha\beta}G(x)+G_0(x)$, so that only the diagonal $\sigma$-independent part survives. This is one of the manifestations of STS where we explicitly see through (\ref{Wcumulant}) that all corrections to the correlation functions that come due to the anharmonic term (proportional to $g$) are $\sigma$ independent.

\section{Integral $I(r,a=0,q,\tau)$}\label{appendix:integral}

In this appendix we analyze and for $\tau>0$ calculate the integral
$I(1,0,q,\tau)$ defined in Eqs.\ (\ref{Integral}) and (\ref{Idim}); it can
be written as
\begin{align}\label{IJconnection}
I(1,0,q,\tau)=\frac{1}{4\pi^2}J\boldsymbol(q(1-\tau),\tau\boldsymbol),
\end{align}
where the integral $J(p,\tau)$ was defined in Eq.~(\ref{40}) in the main text.

The integral of the type (\ref{40}) was studied in Refs.~\onlinecite{Dotsenko+95,Guida+98} in the context of correlation functions in a two-dimensional Ising model.
It can be expressed in terms of the hypergeometric functions, defined as
\begin{align}\label{hyperg-def}
_pF_q(a_1,\ldots,a_p;b_1,\ldots,b_q;z)=\sum_{k=0}^\infty \frac{(a_1)_k\ldots (a_p)_k}{(b_1)_k\ldots(b_q)_k}\frac{z^k}{k!},
\end{align}
where $(a)_k=\Gamma(a+k)/\Gamma(a)$, while $p$ and $q$ are positive integers. In the following we use results from Ref.~\onlinecite{Guida+98} (which are equivalent to those of Ref.~\onlinecite{Dotsenko+95} once one employs some identities between the hypergeometric functions).  It is important to
note that the integral $J(p,\tau)$ diverges for $p=1$ (or more generally integer $p$)  but that the formula below provides an analytic continuation in $p$ away from the poles at integer $p$. In turns, it gives an analytical continuation for the original integral, which
we will call $I_{\textrm{cont}}(1,0,q,\tau)$. We now need to discuss how this continuation is relevant for the physical problem studied here.

Our main objective, as discussed in the main text, is to extract from the poles in $1/\tau^2$ and $1/\tau$ of this integral
the logarithmic behaviors at large distance $r$. How to do that is explained in Appendix \ref{app:poles}. In addition
we will restrict our study to $|q| \leq 1$. As was discussed in the Introduction, the divergences that appears at $q=1$ have important physical consequences and are related to the screening mechanism unveiled by \citet{Toner+90}. To make progress for
$q>1$ one needs to reexamine the entire perturbation expansion, which we defer to a future work. The considerations directly useful for the present work are presented in the first part of this appendix.

Still, it is not devoid of interest to also study (i) the analytical continuation $I_{\textrm{cont}}(1,0,q,\tau)$ for arbitrary $q$ and
(ii) the finite parts of the integral, even if it will not contain the universal information sought here. These considerations are reported in the second part of this appendix.

\subsection{Study useful for $|q| \leq 1$ and the divergent parts}

Using the results from Ref.~\onlinecite{Guida+98}, for the integral (\ref{40}) one obtains
\begin{widetext}
\begin{align}\label{Jguida}
J(p,\tau)=&-\frac{\sin^2(\pi  p)}{2\sin^2(\pi p+2\pi\tau)}
\big\{\left[\cos(2\pi p)+\cos(2\pi p+4\pi\tau)-2\right]B(1-p,1+p)^2 B(1-p,1-2\tau)^2\notag\\ &\times{_3F_2}(1-p,1-2\tau,2-2\tau;2,2-p-2\tau;1)^2\notag\\
&+\left[\cos (4 \pi  \tau )-1\right] B(1+p,2 \tau-1)^2 B(1+p,2\tau)^2
{_3F_2}(p,1+p,2\tau;p+2\tau,1+p+2\tau;1)^2\notag\\
&-4\sin(\pi  p)\sin (2\pi\tau) B(1-p,1+p) B(1-p,1-2\tau) B(1+p,2\tau-1) B(1+p,2\tau)\notag\\
&\times{_3F_2}(1-p,1-2\tau,2-2\tau;2,2-p-2\tau;1)
{_3F_2}(p,1+p,2\tau;p+2\tau,1+p+2\tau;1)
\big\},
\end{align}
where $B(x,y)$ is the beta function. We need a simplified form of Eq.~(\ref{Jguida}) valid at small values of $\tau$.

The excess of the hypergeometric function ${_3F_2}(a,b,c;d,e;z)$ is defined as $d+e-a-b-c$ and for $z=1$ the excess has to be positive in order for ${_3F_2}(a,b,c;d,e;z=1)$ to be  defined by its series (\ref{hyperg-def}). A naive expansion
\begin{align}
{_3F_2}(1-p,1-2\tau,2-2\tau;2,2-p-2\tau;1)={_3F_2}(1-p,1,2;2,2-p;1)+\mathcal{O}(\tau)
\end{align}
of one of the hypergeometric functions of Eq.~(\ref{Jguida}) is not possible. Therefore, although the hypergeometric functions of Eq.~(\ref{Jguida}) are well defined, since they have positive excess $2\tau$, we cannot perform a Taylor expansion around $\tau=0$.

In the following we use particular relations between the hypergeometric functions in order to transform Eq.~(\ref{Jguida}) into a suitable form that can be Taylor expanded at small $\tau$. After transforming ${_3F_2}(1-p,1-2\tau,2-2\tau;2,2-p-2\tau;1)$ of Eq.~(\ref{Jguida}) using the one-term transformation rule\cite{Prudnikov} valid for $a,s>0$
\begin{align}\label{transform}
_3F_2(a,b,c;d,e;1)=\frac{\Gamma(d)\Gamma(e)\Gamma(s)} {\Gamma(a)\Gamma(b+s)\Gamma(c+s)} {_3F_2}(d-a,e-a,s;b+s,c+s;1),
\qquad s=d+e-a-b-c,
\end{align}
one obtains the following result valid at $p<1$,\begin{align}\label{Jguidafinal}
\frac{J(p,\tau)}{4\pi^2}=&-\frac{\pi ^2 p^2 [\cos (2 \pi  p+4 \pi\tau)+\cos(2\pi  p)-2]}{8\sin^2(2\pi \tau)\sin^2(\pi  p+2\pi \tau))} {_3F_2}(1+p,1-2\tau,2\tau;1,2;1)^2\notag\\
&+\frac{\pi ^2 p^3 (2 \tau-1) (p+2 \tau)
\Gamma (2 \tau-1)^2 }{2\sin ^2(\pi  p+2\pi\tau))\Gamma (1-p)^2 \Gamma (p+2 \tau+1)^2}{_3 F_2}(p,1+p,2\tau;p+2\tau,1+p+2\tau;1) {_3 F_2}(1+p,1-2 \tau,2\tau;1,2;1)\notag\\
&+\frac{p^2 (1-2 \tau)^2 \sin^2(2 \pi  \tau) \Gamma (1+p)^2
\Gamma (2 \tau-1)^4\Gamma (1-p-2\tau)^2}{4\pi^2 \Gamma(1-p)^2 \Gamma(1+p+2\tau)^2}{_3F_2}(p,1+p,2\tau;p+2\tau,1+p+2\tau;1)^2.
\end{align}
The expression (\ref{Jguidafinal}) has a positive excess in the hypergeometric functions under the two conditions $\tau>0$ and $p<1$. The main   advantage however of Eq.~(\ref{Jguidafinal}) is  that it is more  convenient  for Taylor expansion around $\tau=0$. We need to expand the two hypergeometric functions of Eq.~(\ref{Jguidafinal}). The first one, $_3 F_2(1+p,1-2\tau,2\tau;1,2;1)$, can be  expanded safely, since it is well-defined in the expansion limit $\tau\to 0$ for $p<1$. One obtains
\begin{align}\label{expansion1}
&_3 F_2(1+p,1-2\tau,2\tau;1,2;1)=1+ 2\left[1-\gamma_{E}-\psi(-p)+\frac{1}{p}\right]\tau+\mathcal{O}(\tau^2).
\end{align}
We employed the Gauss relation $_2F_1(a,b;c;1)=\Gamma(c)\Gamma(c-a-b)/[\Gamma(c-a)\Gamma(c-b)]$ and used the fact that $_3F_2$ reduces to $_2F_1$ for $\tau=0$, see the definition (\ref{hyperg-def}). However, the expansion of the second hypergeometric function of Eq.~(\ref{Jguidafinal}), ${_3F_2}(p,1+p,2\tau;p+2\tau,1+p+2\tau;1)$, can not be done directly, since for $\tau=0$ it has zero excess. To overcome this problem, we use the following relation \cite{functions.wolfram.07.27.17.0018.0-spliting}
\begin{align}
{_3F_2}(p,1+p,2\tau;p+2\tau,1+p+2\tau;1)=&\frac{(1+p)(1+2\tau)}{1+p+2\tau} {_3F_2}(p,1+p,2\tau;p+2\tau,2+p+2\tau;1)\notag\\
&+\frac{p(1-2\tau)}{p+2\tau}{_3F_2}(p,1+p,2\tau;1+p+2\tau,1+p+2\tau;1),
\end{align}
which connects ${_3F_2}(p,1+p,2\tau;p+2\tau,1+p+2\tau;1)$ with two other hypergeometric functions that have an excess of $1+\tau$ and hence are well defined for $\tau=0$. One then obtains
for the expansion\begin{align}\label{expansion2}
&_3 F_2(p,1+p,2\tau;p+2\tau,1+p+2\tau;1)=2+ 2\left[2\gamma_{E}+2\psi(p)+\frac{1}{p}\right]\tau+\mathcal{O}(\tau^2).
\end{align}
Combining Eqs.~(\ref{expansion1}), (\ref{expansion2}), and (\ref{Jguidafinal}), one obtains the expanded form of Eq.~(\ref{Jguida}),
\begin{align}\label{Jexpfinal}
\frac{J(p,\tau)}{4\pi^2}=&\frac{p^2}{8\tau^2}+\frac{p^2}{4\tau} \left[2-2\gamma_{E}-\psi(p)-\psi(-p)\right]+\mathcal{O}(\tau^0).
\end{align}
Finally, using Eqs.~(\ref{IJconnection}) and (\ref{Jexpfinal}), we obtain the result reported in the text (\ref{I(1)}).

It is interesting to notice that Eq.~(\ref{Jguida}) satisfies the relation
$J(p,\tau)=J(p,1/2-\tau)$ when $\tau>0$ and $1/2-\tau>0$, i.e.~for $0<\tau<1/2$. This significantly simplifies the Taylor expansion, since Eq.~(\ref{Jguida}) could  easily and directly be expanded around $\tau=1/2$, because the excess $2\tau$ of the hypergeometric functions is then always positive and hence they are well behaved. Performing the calculation in this way one obtains
\begin{align}
\frac{J(p,\tau)}{4\pi^2}=&\frac{p^2}{8(1/2-\tau)^2}+\frac{p^2}{4(1/2-\tau)} \left[2-2\gamma_{E}-\psi(p)-\psi(-p)\right]+\mathcal{O}\boldsymbol((1/2-\tau)^0\boldsymbol),
\end{align}
which is in agreement with the result (\ref{Jexpfinal}). This observation suggests that a more symmetric form of the expansion  reads
\begin{align}\label{Jfinsym}
\frac{J(p,\tau)}{4\pi^2}=&\frac{p^2}{32\tau^2(1/2-\tau)^2}+\frac{p^2}{8\tau(1/2-\tau)} \left[-2\gamma_{E}-\psi(p)-\psi(-p)\right]+\mathcal{O}\boldsymbol(\tau^0(1/2-\tau)^0\boldsymbol).
\end{align}

\subsection{Further study of the analytical continuation}

We should notice that although Eq.~(\ref{Jexpfinal}) is formally obtained from Eq.~(\ref{Jguidafinal}) that is valid for $p<1$, Eq.~(\ref{Jexpfinal}) is an even function of $p$ [as well as the starting ones of Eqs.~(\ref{40}) and (\ref{Jguida})] and therefore Eq.~(\ref{Jexpfinal}) holds at any $p$ by the parity transformation. Hence we can define the integral analytically continued to the whole complex plane:
\begin{align}\label{I(1) final}
I_{\textrm{cont}}(1,0,q,\tau)=\frac{q^2}{8\tau^2}+\frac{q^2 \left[1-2\gamma_{E}-\psi(q)-\psi(-q)\right]}{4\tau}+\mathcal{O}(\tau^0) \quad\text{for}\quad q\notin\text{integers}.
\end{align}
For positive integer $q=N$, one can not expand the function $\psi\boldsymbol(-q(1-\tau)\boldsymbol)$ from Eq.~(\ref{Jexpfinal}) at small $\tau$ due to the divergence of $\psi(-N)$. In Fig.~\ref{fig2} we show a comparison between the exact result and its small $\tau$ expanded form for the $1/\tau$ divergent part of $I(1,0,q,\tau)$.

\begin{figure}
\includegraphics[width=0.6\columnwidth]{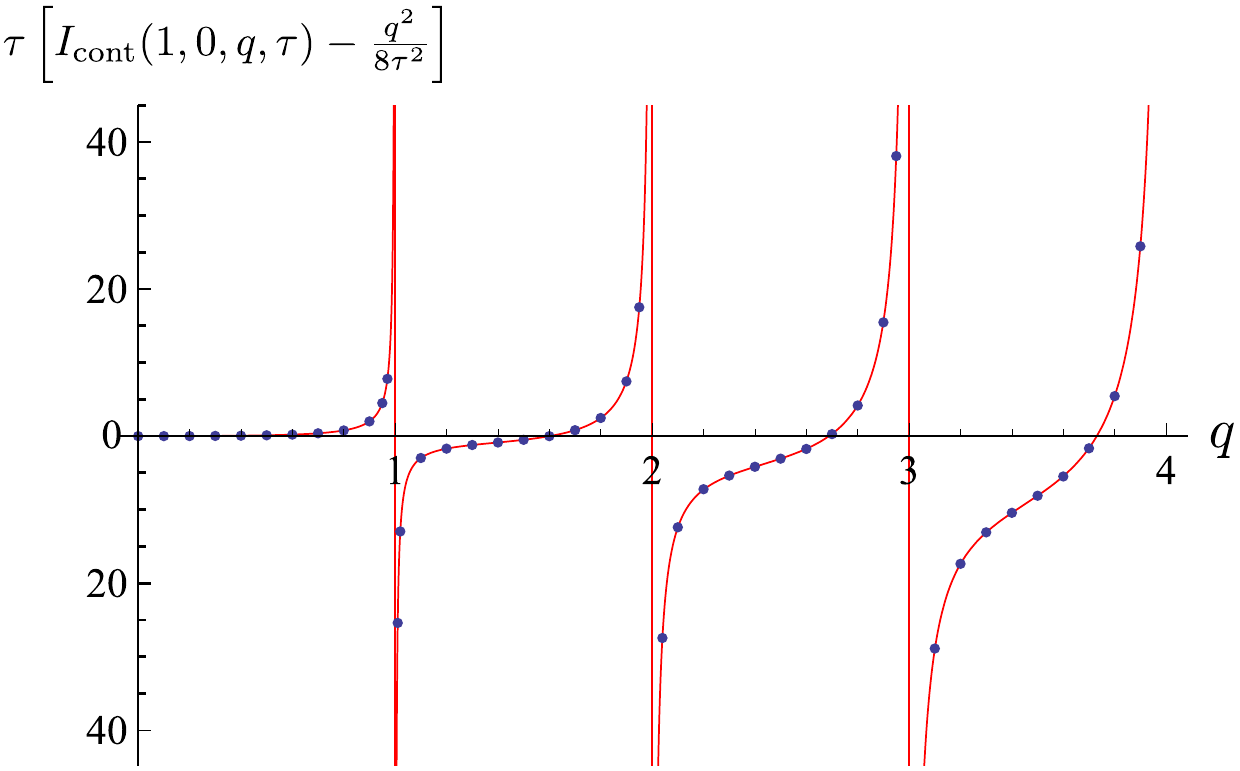}
\caption{Numerical evaluation of $1/\tau$ divergent part of $I_{\textrm{cont}}(1,0,q,\tau)$ for $\tau=10^{-5}$. Solid line represents the data obtained using Eq.~(\ref{I(1) final}), while the dots are obtained using the exact expression (\ref{Jguida}) plugged in into Eq.~(\ref{IJconnection}). The plotted function is even in $q$ and we only show the region $q>0$.}\label{fig2}
\end{figure}

We finally notice that Eq.~(\ref{Jexpfinal}) diverges, once $p$ approaches an integer. Due to the parity of Eq.~(\ref{Jexpfinal}) with respect to $p$, we can restrict to the region $p>0$. It is then convenient to rewrite $\psi(p)+\psi(-p)=2\psi(p)+\pi\cot(\pi p)+1/p$. As a result, the divergence arises due to the cotangent term. Therefore, the most divergent part of Eq.~(\ref{Jexpfinal}) at $\tau\to 0$ and $p\to N$ with $N$ being a positive integer, is
\begin{align}\label{Jspec}
\frac{J(p,\tau)}{4\pi^2}=\frac{p^2}{8\tau^2}-\frac{p^2}{4\tau}\pi\cot(\pi p)+R(p),
\end{align}
where $R(p)$ is  the $\tau$-independent part in Eq.~(\ref{Jexpfinal}), contained in $\mathcal{O}(\tau^0)$. Direct expansion of Eq.~(\ref{Jguida}) shows that it has a divergence of the type $N^2/[4(p-N)^2]$ for any integer $N$, so $R(p)$ can be rewritten as $R(p)=\sum_{N=1}^\infty \frac{p^2}{4(p-N)^2}+\ldots=\frac{p^2}{4}\psi^{(1)}(1-p)+\ldots$, taking into account only the most divergent terms $\mathcal{O}[(q-N)^{-2}]$, but not the terms that diverge as $\mathcal{O}[(q-N)^{-1}]$. Here $\ldots$ denotes other subleading divergent terms when $q\to N$. Therefore,
\begin{align}\label{JguidaNexpansion}
\frac{J(N+h,\tau)}{4\pi^2}=&\frac{(N+h)^2}{8\tau^2}-\frac{(N+h)^2}{4\tau}\left[\frac{1}{h} +\mathcal{O}(h)\right]+\frac{N^2}{4h^2}\left[1+\mathcal{O}(h/N)\right]\notag\\
=&\frac{N^2}{8\tau^2}\left[1+\mathcal{O}(h/N)\right]-\frac{N^2}{4\tau h}\left[1+\mathcal{O}(h/N)\right]+\frac{N^2}{4h^2}\left[1+\mathcal{O}(h/N)\right].
\end{align}
Using Eqs.~(\ref{IJconnection}) and (\ref{JguidaNexpansion}) we recognize $h=-N\tau$, and therefore
\begin{align}
I_{\textrm{cont}}(1,0,N,\tau)=\frac{N^2+2N+2}{8\tau^2}+\mathcal{O}(\tau^{-1}).
\end{align}
In Fig.~\ref{fig3} we show a numerical comparison between the obtained results.

\begin{figure}
\includegraphics[width=0.6\columnwidth]{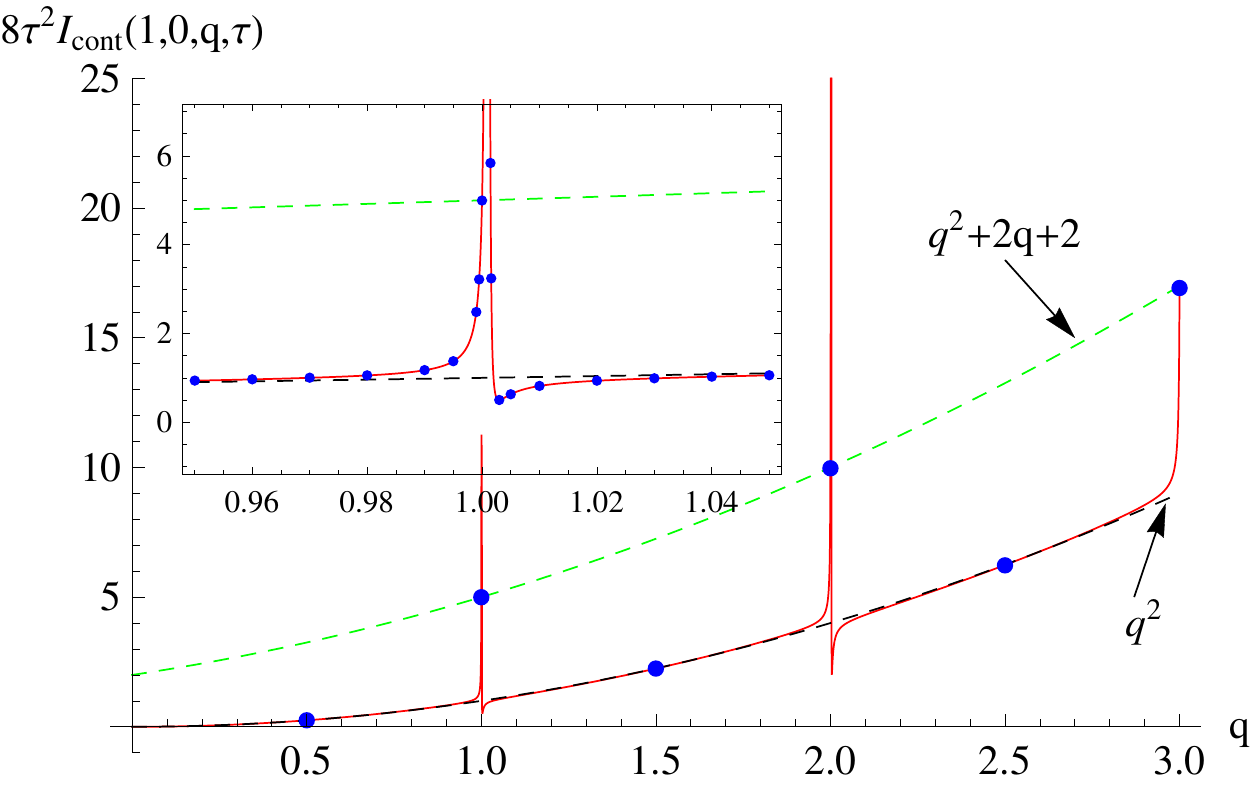}
\caption{Numerical evaluation of $1/\tau^2$ divergent part of $I_{\textrm{cont}}(1,0,q,\tau)$ for $\tau=10^{-3}$. For non-integer $q$ values, $8\tau^2 I_{\textrm{cont}}(1,0,q,\tau)$ equals $q^2$, while for integer $q$ it is $q^2+2q+2$, as indicated by the dashed lines. The dots represent values obtained using the exact expression (\ref{Jguida}) plugged in into Eq.~(\ref{IJconnection}) at particular values $q=0.5,1,1.5,2,2.5,3$. By solid line we show the result  (\ref{Jspec}) plugged in into Eq.~(\ref{IJconnection}). The inset shows the zoomed result around $q=1$, and we notice a very good agreement between the results (\ref{Jguida}) and (\ref{Jspec}). For special $q$ values such that $p=q(1-\tau)$ approaches unity or any other integer, we notice the divergence of $\tau^2I_{\textrm{cont}}(1,0,q,\tau)$.}\label{fig3}
\end{figure}

In order to get the $\mathcal{O}(\tau^{-1})$ term in the previous formula, one should evaluate the $\mathcal{O}(\tau^2)$ terms in (\ref{expansion1}) and (\ref{expansion2}) that we have not succeeded. However, we have verified the following two results:
\begin{gather}
I_{\textrm{cont}}(1,0,1,\tau)=\frac{5}{8\tau^2}-\frac{1}{2\tau}+\mathcal{O}(\tau^0),\\
I_{\textrm{cont}}(1,0,2,\tau)=\frac{5}{4\tau^2}-\frac{4}{\tau}+\mathcal{O}(\tau^0).
\end{gather}

\section{Integral $I(r,a,q,\tau=0)$}\label{appendix:finite a integrals}

In this appendix we consider the integral (\ref{Integral}) from the main text  at $\tau=0$, i.e.,
\begin{align}\label{Integral1}
I(r,a,q,\tau=0)=&\frac{1}{4\pi^2}\int\dif^2 x\dif^2 y\frac{1}{\left[a^2+(\mathbf{x}-\mathbf{y})^2\right]^{2}} \left\{\left[\frac{(\mathbf{x}-\mathbf{r})^2+a^2}{(\mathbf{y}-\mathbf{r})^2+a^2}\  \frac{\mathbf{y}^2+a^2}{\mathbf{x}^2+a^2}\right]^{q}-1\right\},
\end{align}
and evaluate it for $q=1$ and in the limit $q\to 0$. We notice the invariance $I(r,a,q,0)=I(r/a,1,q,0)$ and therefore we set $a=1$, recovering $a$ in the final result.

\subsection{The case $I(r,a,q=1,\tau=0)$}\label{I(q=1)}

We start with the integral (\ref{Integral1}) in the case $q=1$. We represent the two-dimensional vectors $\mathbf{x}$ and $\mathbf{y}$ in polar coordinates. After performing the two angular integrations, and then the integration over $x$, the result can be expressed in the following form
\begin{gather}\label{J1}
I(r,1,1,0)=\int_0^\infty \dif y [j_1(r,y)+j_2(y)+j_3(r,y)],
\end{gather}
where
\begin{gather}
j_1(r,y)=\frac{1}{2y(y^2+4)}\left[2-y^2+\frac{(y^2+1)(y^2-2r^2-2)} {\sqrt{r^4-2r^2(y^2-1)+(y^2+1)^2}}\right],\\
j_2(y)=-\frac{2}{(y^2+4)^{3/2}}\mathrm{arctanh}\left(\frac{y\sqrt{y^2+4}}{y^2+2}\right),\\
\label{j3def}
j_3(r,y)=\frac{1}{y^2(y^2+4)^{3/2}} \left[\frac{(y^2+1)(r^2y^2+2y^2+2r^2+2)}{\sqrt{r^4-2r^2(y^2-1)+(y^2+1)^2}}-2\right] \mathrm{arctanh}\left(\frac{y\sqrt{y^2+4}}{y^2+2}\right).
\end{gather}
After introducing a new variable $t=y^2+2$, one easily performs the integration of $j_1(r,y)$ and obtains
\begin{align}\label{j1int}
\int_0^\infty \dif y j_1(r,y)=&-\frac{1}{4}\ln\left(\frac{r^2+1}{8}\right)-\frac{3r^2+9}{8\sqrt{(r^2+1)(r^2+9)}} \ln\left[\frac{(r^2+5)\sqrt{(r^2+1)(r^2+9)}+r^4+10r^2+17}{8}\right]
\notag\\
=&-2\ln(r)+\frac{3}{2}\ln2+\mathcal{O}\left(\frac{1}{r^2}\right).
\end{align}
The function $j_2(y)$ is $r$-independent and gives a constant contribution
\begin{align}\label{j2int}
\int_0^\infty \dif y j_2(y)=-\ln2.
\end{align}
The remaining yet unevaluated integral is $\int_0^\infty\dif y\, j_3(r,y)$. We notice that it contains a $r$-independent part [the term $-2$ from the square brackets in Eq.~(\ref{j3def})], which is necessary to make $j_3(r,y)$ nonsingular at $y\to 0$. From the square root of the denominator of Eq.~(\ref{j3def}), we conclude that $j_3(r,y)$ is sharply peaked when $y$ takes  values around $r$, with the height of the peak equal to $\ln(r)$ in the limit of large $r$. Therefore, we split $j_3(r,y)$ in the following form
\begin{gather}\label{j3}
j_3(r,y)=j_{31}(r,y)+j_{32}(r,y),
\end{gather}
where
\begin{gather}\label{j31}
j_{31}(r,y)=\frac{r^2 \mathrm{arctanh}\left(\frac{y\sqrt{y^2+4}}{y^2+2}\right)} {y\sqrt{r^4-2r^2(y^2-1)+(y^2+1)^2}}.
\end{gather}
In the last equation we keep $y$ in the denominator in order make $j_{32}(r,y)$ decaying faster than $1/y$ at large $y$. We emphasize here that the special choice (\ref{j31}) for the first term of Eq.~(\ref{j3}) is not unique. For example, another possibility would be $\frac{2r^2\ln(y+1)}{y\sqrt{r^4-2r^2(y^2-1)+(y^2+1)^2}}$, which is equally good, as it is integrable and  describes well the function $j_3(r,y)$ around its maximum. For simplicity, in the following we use the choice given by Eq.~(\ref{j31}).

The subleading part of $j_3(r,y)$ contained in $j_{32}(r,y)=j_3(r,y)-j_{31}(r,y)$ can be safely expanded at large $r$, as its main contribution in the integral comes from small values of $y$, i.e., from the region $y \ll r$. Using Eqs.~(\ref{j3def}), (\ref{j3}), and (\ref{j31}) one  obtains
\begin{gather}
j_{32}(r,y)=\left[\frac{y^2+3}{(y^2+4)^{3/2}}-\frac{1}{y}\right] \mathrm{arctanh}\left(\frac{y\sqrt{y^2+4}}{y^2+2}\right)+\mathcal{O}\left(\frac{1}{r^2}\right);
\end{gather}
therefore
\begin{gather}\label{j32}
\int_0^\infty \dif y j_{32}(r,y)=-\frac{\pi^2}{6}-\frac{\ln 2}{2} +\mathcal{O}\left(\frac{1}{r^2}\right).
\end{gather}
We perform the remaining integration $\int_0^\infty \dif y\, j_{31}(r,y)$  in the following way: We first calculate
\begin{align}\label{J31full}
J_{31}(r,y)=&\int \dif r \frac{j_{31}(r,y)}{r}=\frac{\mathrm{arctanh}\left(\frac{y\sqrt{y^2+4}}{y^2+2}\right)}{2y} \ln\left(1+r^2-y^2+\sqrt{r^4-2r^2(y^2-1)+(y^2+1)^2}\right).
\end{align}
Then, we expand $J_{31}(r,y)$ at large $r$ at lowest order in $r$ and get \begin{gather}\label{J31}
J_{31}^{\mathrm{a}}(r,y)=\frac{\mathrm{arctanh}\left(\frac{y\sqrt{y^2+4}}{y^2+2}\right)}{y}\ln r+\ldots,
\end{gather}
where by $\ldots$ we denote the subleading terms at large $r$. The expansion of Eq.~(\ref{J31}) is a good approximation of $J_{31}(r,y)$ of Eq.~(\ref{J31full}) only for $y\lesssim r$. For $y>r$, the function $J_{31}(r,y)$  sharply drops to zero, contrary to its expanded form (\ref{J31}), as one can see from the expansion of $J_{31}(r,y)$. In the vicinity of $r$, at leading order one obtains $J_{31}(r,y=r+\delta)=(\ln^2r/r)[1-\delta/\ln r]$ for $\delta\ll 1$. Such an expansion determines a very large slope $\mathcal{O}(1/\ln r)$ for the deviation of the function $J_{31}(r,y)$ around the point $y=r$. Therefore, we integrate the expanded result (\ref{J31}) over $y$ in the interval $[0,r]$ and get
\begin{align}\label{J31result}
\int_0^r \dif y J_{31}^{\mathrm{a}}(r,y)=&-\text{Li}_2\left[-\frac{1}{2} r
   \left(r+\sqrt{r^2+4}\right)\right]
   \ln(r)-\frac{1}{4} \ln(r) \ln
   ^2\left(\frac{2}{r^2+r\sqrt{r^2+4}
   +2}\right)+\ldots\notag\\
   =&\ln^3 r+\frac{\pi^2}{6}\ln r+\mathcal{O}\left(\frac{1}{r^2}\right)+\ldots
\end{align}
The subleading terms of Eq.~(\ref{J31result}), denoted by $\ldots$, originate from the subleading terms of Eq.~(\ref{J31}) which do not change the stated result in Eq.~(\ref{J31result}), as one can check, e.g., numerically. Finally, assuming the following form $\int_0^\infty \dif y\, j_{31}(r,y)=A\ln^2 r+ B\ln r+C+\ldots$, one obtains
\begin{align}
\int\dif r\int_0^\infty\dif y \frac{j_{31}(r,y)}{r}=\frac{A}{3}\ln^3 r+\frac{B}{2}\ln^2 r+C\ln r+\ldots.
\end{align}
After comparing the last expression with Eq.~(\ref{J31result}) one gets $A=3,B=0$, and $C=\pi^2/6$. Combining this result with Eqs.~(\ref{j3}) and (\ref{j32}) one obtains
\begin{align}\label{j3int}
\int_0^\infty\dif y j_3(r,y)=3\ln^2 r-\frac{\ln2}{2}+\mathcal{O}\left(\frac{1}{r^2}\right).
\end{align}
Collecting the obtained results (\ref{j1int}), (\ref{j2int}), and (\ref{j3int}) and using (\ref{J1}), one obtains
\begin{align}
I(r,a,1,0)=3\ln^2\left(\frac{r}{a}\right)-2\ln\left(\frac{r}{a}\right) +\mathcal{O}\left(\frac{a^2}{r^2}\right),
\end{align}
where we have recovered the parameter $a$.

\subsection{The case $I(r,a,q\to 0,\tau=0)$}

The leading order term at $q\to 0$ of Eq.~(\ref{expansion}) is given by $I_1(r)$ of Eq.~(\ref{I1}). We could evaluate it using a procedure  similar to the one employed during the evaluation of $I(r,a,1,0)$, see Appendix~\ref{I(q=1)}. First, one  performs the two angular integrations, where one employs the following non-elementary integral
\begin{align}
\int_0^{2\pi}\dif\theta\ln (c-\cos\theta)=-2\pi\ln\left(2c-2\sqrt{c^2-1}\right),
\end{align}
which can, e.g.,  be obtained by making use of the Jensen formula.\cite{Ahlfors}
After performing one spatial integration, one obtains the result of the form
\begin{align}
I_1(r)=\int_0^\infty\dif y \left[\frac{y^2+2}{\sqrt{y^2+4}} \mathrm{arctanh}\left(\frac{y\sqrt{y^2+4}}{y^2+2}\right)-y\ln(y^2+1)\right] \ln\left(\frac{r^2+y^2+1+\sqrt{r^4-2r^2(y^2-1)+(y^2+1)^2}}{2(y^2+1)}\right).
\end{align}
\end{widetext}
For the last integral one could use a  procedure similar to the one employed above for the integral (\ref{J1}). The final result reads
\begin{align}
I_1(r)=\ln^2(r)+\ln (r)+ \mathcal{O}\left(r^0\right),
\end{align}
and therefore
\begin{align}
I(r,a,q,0)=q^2 \left[\ln^2\left(\frac{r}{a}\right)+\ln\left(\frac{r}{a}\right) +\mathcal{O}\left(\frac{r^0}{a^0}\right)\right]+\mathcal{O}(q^4).
\end{align}

\section{Connection between the finite-$a$ and the dimensional method}\label{appendix:connection}

\label{app:connection}

\subsection{From poles to logarithms}
\label{app:poles}

Let us define $\tilde I(\tilde r,\tau) := \tilde r^{4 \tau} I(r,a,q,\tau)$ and $\tilde r=r/a$ [compare with the expression for $W_2$ of Eq.~(\ref{W2int})]. Hence the limit
$a \to 0$ is the same as $\tilde r \to \infty$. The dimensional
method gives, at fixed $\tau>0$:
\begin{eqnarray} \label{dim1}
\tilde I(\tilde r,\tau)  \simeq_{\tilde r \to \infty} \tilde r^{4 \tau} \left[\frac{b_2(q)}{\tau^2} + \frac{b_1(q)}{\tau} + \mathcal{O}(1)\right],
\end{eqnarray}
where $b_2(q)$ and $b_1(q)$ are even functions of $q$ with a regular Taylor expansion in $q$ at $q=0$.
On the other hand, from the finite-$a$ method, we know that
\begin{eqnarray} \label{dim2}
&& \tilde I(\tilde r,\tau=0) \simeq_{\tilde r \to \infty}  B_2(q) \ln^2 \tilde r + B_1(q) \ln \tilde r + \mathcal{O}(1).
\end{eqnarray}
To match the two, we first observe that the integral
\begin{eqnarray}
(\tilde r \partial_{\tilde r})^2 \tilde I(\tilde r,\tau)
\end{eqnarray}
is not divergent and can be calculated at $\tau=0$, giving $2 B_2(q)$ from Eq.~(\ref{dim2}), or directly
calculated from Eq.~(\ref{dim1}) in the limit $\tau\to0$ is gives $16 b_2(q)$. Identifying
the two coefficients, one gets
\be
B_2(q) = 8 b_2(q),
\ee
which was used in the main text.

When $b_2(q)$, $B_2(q)$ are non zero one cannot get in full generality a universal result for $B_1(q),b_1(q)$. It is easy to see since by
simply changing the cutoff $a$ by a finite scale $B_1(q)$ changes. However in the present case we can use the extra parameter, $q$, and
we note that $b_2(q) = q^2/8$. Hence the integral
\be
\hat I(\tilde r,\tau) = \left(1 - \frac{1}{2} q^2 \partial_q^2\right)|_{q=0} \tilde I(\tilde r,\tau)  =\tilde r^{4\tau} \frac{\tilde b_1(q)}{\tau} + \mathcal{O}(1),
\ee
where $\tilde b_1(q)=\left(1-\frac{1}{2} q^2 \partial_q^2\right)|_{q=0} b_1(q)$ has a Taylor expansion starting at $\mathcal{O}(q^4)$. In the finite-$a$ method this new integral can have only a logarithmic divergence at large $\tilde r$, namely:
\bea
\hat I(\tilde r,\tau) = \tilde B_1(q) \ln \tilde r + \mathcal{O}(1),
\eea
and applying the same reasoning as above to the finite integral $\tilde r \partial_{\tilde r}  \hat I (\tilde r,\tau)$ we obtain
\bea
\tilde B_1(q)=4 \tilde b_1(q).
\eea

Hence in conclusion if $b_2(q)=q^2/8$ we conclude that:
\begin{align}
\tilde I(\tilde r,\tau=0) \simeq&_{\tilde r \to \infty}\  q^2 [ \ln^2 \tilde r + \gamma (\ln \tilde r) ] \\
& + 4 \tilde b_1(q) \ln \tilde r + \mathcal{O}(1),
\end{align}
where $\gamma$ is non-universal,
as used in the text.

\subsection{An illustrative example}
\label{app:connection2}

In this appendix we study an example the connection between the two  approaches used, the first one which keeps a finite ultraviolet cutoff $a$ but uses $\tau=0$, and the second dimensional method where $a=0$ but $\tau>0$. In order to understand the connection, let us consider the following example:
\begin{align}\label{Jint}
K(r,a,\tau)=\left(\frac{r}{a}\right)^{2\tau} r^{-2\tau}\int_{|x|<L}\dif^2 x \frac{1}{\left(x^2+a^2\right)^{1-\tau}},
\end{align}
which, although being simple, resembles the structure of the terms we encountered in Eqs.~(\ref{W2int}) and (\ref{Integral}) and gives the essence of the difference of the two methods. The exact evaluation of Eq.~(\ref{Jint}) is elementary and one obtains
\begin{align}\label{Jresult}
K(r,a,\tau)=2\pi \left(\frac{r}{a}\right)^{2\tau} r^{-2\tau} \frac{(L^2+a^2)^\tau-a^{2\tau}}{2\tau}.
\end{align}
Now we evaluate Eq.~(\ref{Jint}) using the finite-$a$ method. There, one first expands the integrand at small $\tau$, followed by the integration. Doing these two steps, one obtains
\begin{align}\label{Ja}
K^{a}(r,a,\tau)=&\int_{|x|<L}\dif^2 x \frac{1}{x^2+a^2}+\mathcal{O}(\tau)\notag\\
=&2\pi\ln\left(1+\frac{L^2}{a^2}\right)+\mathcal{O}(\tau).
\end{align}
In the dimensional method, one first expands the integrand at small $a$, and then performs the integration. For $\tau>0$, this procedure leads to a convergent integral, since the short-distance divergence is avoided by keeping $\tau$ finite. In such a way one obtains
\begin{align}\label{Jdim}
K^{\mathrm{dim}}(r,a,\tau)=&\left(\frac{r}{a}\right)^{2\tau} r^{-2\tau}\left[\int_{|x|<L}\dif^2 x \frac{1}{\left(x^2\right)^{1-\tau}}+\mathcal{O}(a^2)\right]\notag\\
=&2\pi \left(\frac{r}{a}\right)^{2\tau} r^{-2\tau} \left[\frac{L^{2\tau}}{2\tau}+\mathcal{O}(a^2)\right].
\end{align}
In the final step of the dimensional method one expands at small $\tau$ the  result (\ref{Jdim}), finding
\begin{align}\label{Jdimf}
K^{\mathrm{dim}}(r,a,\tau)=2\pi\left[\ln\frac{L}{a}+\frac{1}{2\tau} +\mathcal{O}(\tau)\right]+\mathcal{O}(a^{2-2\tau}).
\end{align}
The obvious discrepancy, at small $a$, between the two results of Eqs.~(\ref{Ja}) and (\ref{Jdimf}), is contained in the term $\pi/\tau$ in the latter; it arises due to setting $a=0$ in the integrand of Eq.~(\ref{Jdim}), which is reminiscent of setting $a$ to zero in the last term of the exact result (\ref{Jresult}). As a consequence, the divergence at $\tau\to 0$ in Eq.~(\ref{Jdimf}) remains, which would have been canceled had we kept a finite $a$ in the dimensional method. Therefore, in order to compare the two results of the finite-$a$ and the dimensional method, one must neglect all the terms that are divergent in the limit $\tau\to 0$ in the final result of the dimensional method.

We note finally that  expressions of the form $a^\tau$, which appear in the integral (\ref{Jint}) and in similar ones, are treated differently in the finite-$a$ method and in the dimensional one. In the former one encounters the limit $\tau=0$ and therefore $a^\tau=1$. In the latter one considers the limit $a\to 0$, and therefore $a^\tau=0$. This difference must taken into account if one wants to compare the results obtained by the two methods.

\section{From moments to distribution}

In principle the knowledge of the cumulants (\ref{cum1}) of the relative phase displacements $\Theta=\theta(r) - \theta(0)$ allows to
learn about the probability distribution function (PDF) of $\Theta$, which could be measured directly, e.g., in numerical simulations.

The presence of poles in $\eta(q)$ at $q=1$, and presumably other integers, as well as the (related) mechanism of screening
discussed in the introduction precluded us for now to have a complete knowledge of ${\cal A}(q)$ and $\eta(q)$ for all $q$,
hence we are not able to fully characterize the PDF.

In a more modest attempt, hopefully avoiding some of these problems, let us focus on the correlation at imaginary $q=- i p$, for which our
result reads:
\bea \label{real}
&& \overline{\langle e^{ p \Theta } \rangle} \simeq \exp\left({ p^2 \tau^2 \ell^2 + [ c p^2- g(p) \tau^2] \ell }\right), \\
&& g(p) = p^2 [2 \gamma_E + \psi(ip) + \psi(-ip)],
\eea
 where $\ell=\ln(r/a)$ and $g(p)$ is a positive even function, increasing for $p>0$, which behaves as $\mathcal{O}(p^4)$ at small $p$ and
as $2 p^2 \ln p$ at large $p$. Note that there are no poles on the real $p$ axis. At this stage (\ref{real}) is just the
generating function of all cumulants of $\Theta$ and the explicit form (\ref{real}) provides a resummation of the Taylor series
in the vicinity of $p=0$. It is not obvious whether this equality holds more globally, i.e., whether additional non-perturbative terms
are also present, as is the case along real $q$. However it is maybe less likely for real $p$.

To test that we must verify that (\ref{real}) is first of all an increasing function of $p$. Clearly, since
$g(p) \simeq 2 p^2 \ln p$ this fails at fixed $\ell$ for some large enough $p$. So this formula cannot
extend to $\tau^2 \ln p > c \ell + \tau^2 \ell^2$. It is probable that for such large $p$ one leaves the
domain of validity of the (renormalized) perturbative expansion and a different calculation must be
performed (such as an instanton calculation for real $p$).

Let us point out an interesting interpretation. Let us rewrite, to the same accuracy at large $\ell$:
\bea
&& \overline{\langle e^{ p \Theta } \rangle} \simeq \exp\left({ c p^2 \ell + p^2 \tau^2 \ell^2 e^{-h(p)/\ell}  }\right)
\eea
where $h(p) =g(p)/p^2 \sim \mathcal{O}(p^2)$ at small $p$ and $h(p) \sim 2 \ln p$ at large $p$. From the
large $p \gg1$ behavior we can rewrite:
\bea
&& \overline{\langle e^{ p \Theta } \rangle} \simeq_{p \gg 1} \exp\left({ c p^2 \ell +  \tau^2 \ell^2 p^{2 - \frac{2}{\ell}}  }\right),
\eea
i.e., indicating some deviations from the Gaussian behavior.

\section{A toy model}
\label{app:toy}

\begin{figure}
\includegraphics[width=0.9\columnwidth]{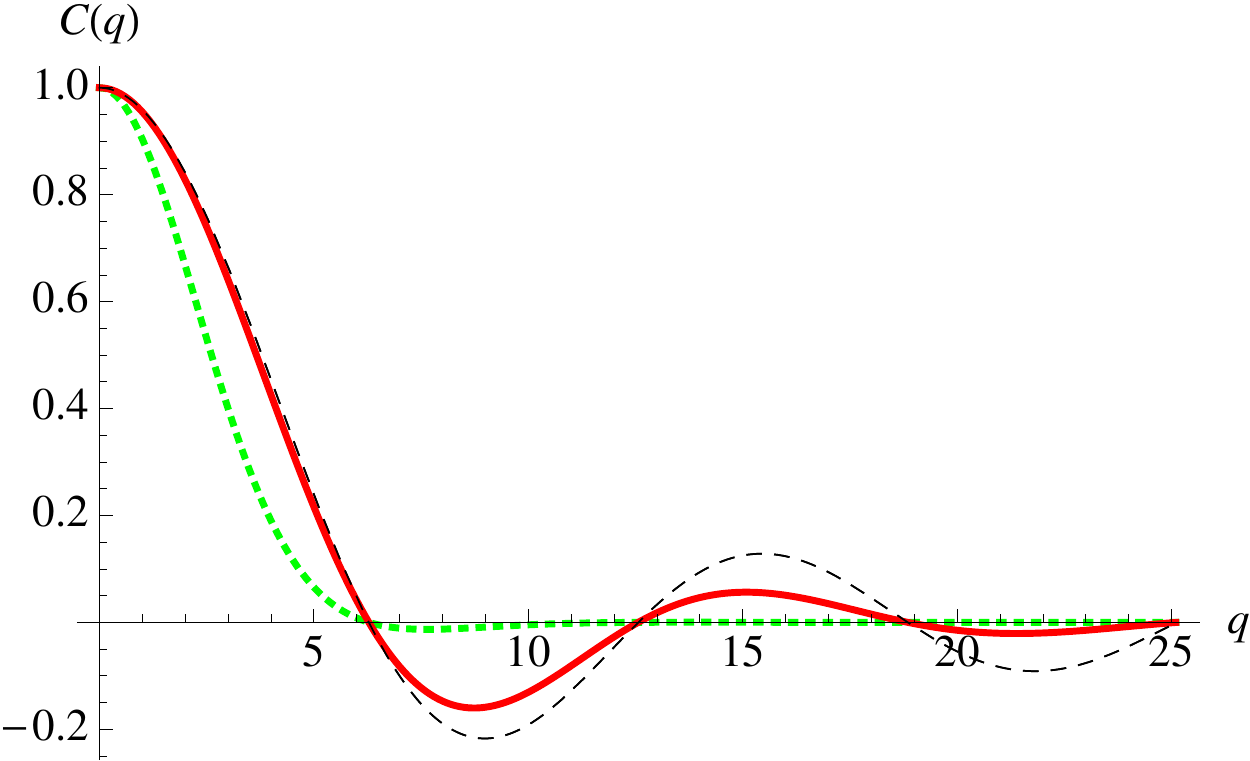}
\caption{The correlation function $C(q)$ for $T=1/5$ (thick, green, dotted), $T=1/20$ (red, thick, solid), and in the limit of $T\to 0$ (black, thin, dashed). In the latter case it approaches $C(q)|_{{T=0}}=2\sin(q/2)/q$.}
\label{f:toy}
\end{figure}\begin{figure}
\includegraphics[width=0.9\columnwidth]{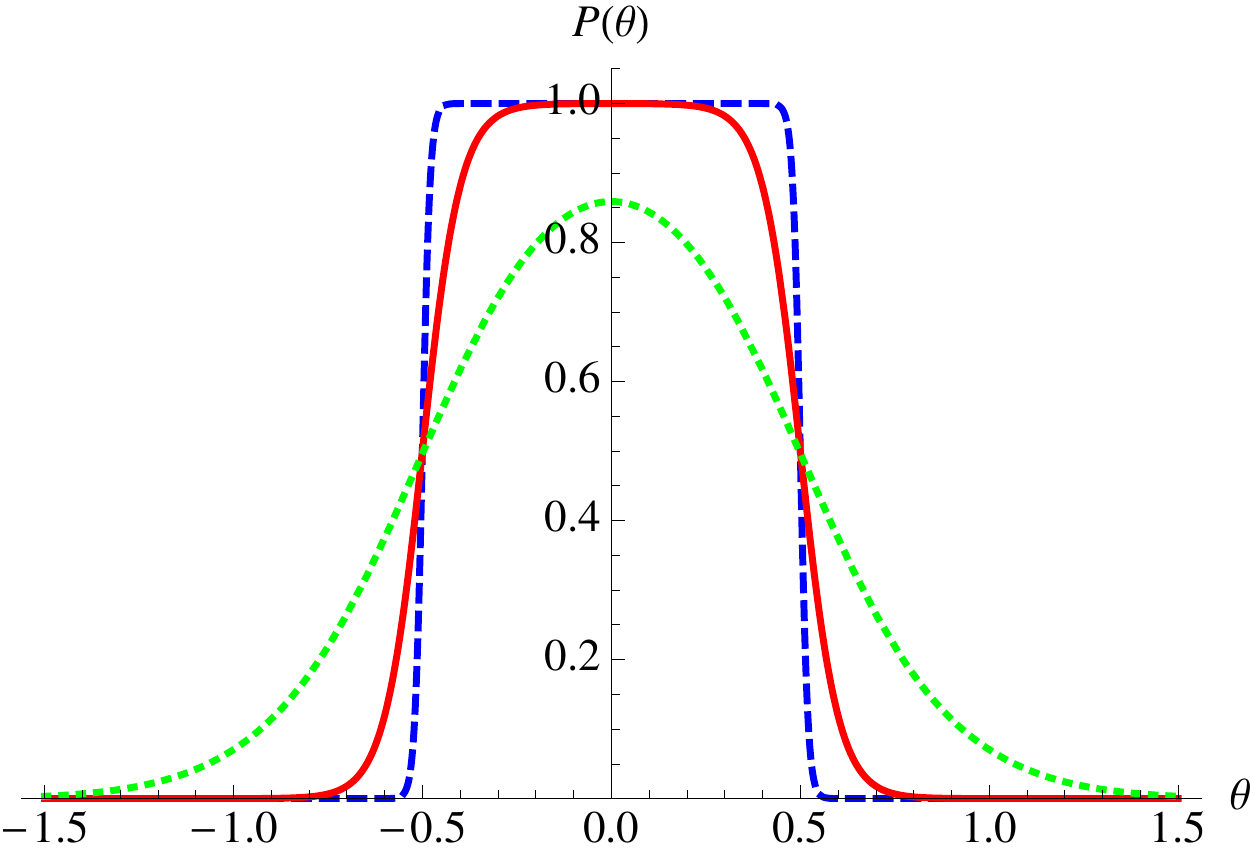}
\caption{The probability distribution (\ref{Ptoy}), for $T=1/100$ (blue, dashed), $T=1/20$ (red, solid), and $T=1/5$ (green, dotted).}
\label{f:Poftheta}
\end{figure}

Here we give results for a  toy model that shows oscillatory behavior in $C(q)$ as a function of $q$.
We write the energy for a particle in a parabola plus a cosine, which has a random shift $\xi\in [-1/2,1/2]$,
\be
{\cal H}[\theta]= \frac12 \theta^{2} - g_{0} \cos(2\pi[\theta+\xi])\ .
\ee
For simplicity we consider the limit of $g_{0}\to \infty$. This restricts $\theta$ to $-\xi$ plus an integer $m$. The expectation of $e^{i q \theta}$, given $\xi$ can then be written as
\bea\label{G2}
\left\langle e^{i q \theta}\right\rangle \Big|_{\xi} &=& \frac{\sum_{m=-\infty}^{\infty} e^{i q (m-\xi)- \frac{(m-\xi)^{2}}{2T} }}{\sum_{m=-\infty}^{\infty} e^{- \frac{(m-\xi)^{2}}{2T}}}\nn\\
&=&\frac{e^{-\frac{q^2 T}{2}} \vartheta _3\left(\pi
   (\zeta -i q T),e^{-2 \pi ^2
   T}\right)}{\vartheta _3\left(\pi  \zeta ,e^{-2
   \pi ^2 T}\right)}\ .\qquad \\
  &&\nn
\eea
where $\vartheta _3$ is the elliptic $\theta$ function and $\zeta=-\xi$.
Then
\be
C(q) = \overline{\left< e^{i q \theta}\right>} = \int_{-1/2}^{1/2} d \xi\, \left< e^{i q \theta}\right>\Big|_{\xi}\ .
\ee
We have plotted the result in Fig.~\ref{f:toy}.
For $T\to 0$, we can restrict the sum in Eq.~(\ref{G2}) to $m=0$. This yields
\be
C(q)\Big|_{T=0} = \int_{-1/2}^{1/2}d \xi\, e^{i q \xi}= \frac2q \sin\left( \frac q2\right)
\ee
The corresponding probability distribution is
\be
P(\theta)\Big|_{T=0} = \Theta\left(-\frac12 < \theta <\frac12   \right)\ .
\ee
This means that $\theta$ is uniformly distributed between $-1/2$ and $1/2$. For higher temperatures, the distribution will be smeared out. It then reads
\be\label{Ptoy}
P(\theta)=\frac{e^{-\frac{\theta ^2}{2 T}}}{\sqrt{2 \pi }
   \sqrt{T} \vartheta _3\left(\pi  \theta ,e^{-2
   \pi ^2 T}\right)}
\ee
This is plotted in Fig.~\ref{f:Poftheta}.


\begin{thebibliography}{42}%
\makeatletter
\providecommand \@ifxundefined [1]{%
 \@ifx{#1\undefined}
}%
\providecommand \@ifnum [1]{%
 \ifnum #1\expandafter \@firstoftwo
 \else \expandafter \@secondoftwo
 \fi
}%
\providecommand \@ifx [1]{%
 \ifx #1\expandafter \@firstoftwo
 \else \expandafter \@secondoftwo
 \fi
}%
\providecommand \natexlab [1]{#1}%
\providecommand \enquote  [1]{``#1''}%
\providecommand \bibnamefont  [1]{#1}%
\providecommand \bibfnamefont [1]{#1}%
\providecommand \citenamefont [1]{#1}%
\providecommand \href@noop [0]{\@secondoftwo}%
\providecommand \href [0]{\begingroup \@sanitize@url \@href}%
\providecommand \@href[1]{\@@startlink{#1}\@@href}%
\providecommand \@@href[1]{\endgroup#1\@@endlink}%
\providecommand \@sanitize@url [0]{\catcode `\\12\catcode `\$12\catcode
  `\&12\catcode `\#12\catcode `\^12\catcode `\_12\catcode `\%12\relax}%
\providecommand \@@startlink[1]{}%
\providecommand \@@endlink[0]{}%
\providecommand \url  [0]{\begingroup\@sanitize@url \@url }%
\providecommand \@url [1]{\endgroup\@href {#1}{\urlprefix }}%
\providecommand \urlprefix  [0]{URL }%
\providecommand \Eprint [0]{\href }%
\providecommand \doibase [0]{http://dx.doi.org/}%
\providecommand \selectlanguage [0]{\@gobble}%
\providecommand \bibinfo  [0]{\@secondoftwo}%
\providecommand \bibfield  [0]{\@secondoftwo}%
\providecommand \translation [1]{[#1]}%
\providecommand \BibitemOpen [0]{}%
\providecommand \bibitemStop [0]{}%
\providecommand \bibitemNoStop [0]{.\EOS\space}%
\providecommand \EOS [0]{\spacefactor3000\relax}%
\providecommand \BibitemShut  [1]{\csname bibitem#1\endcsname}%
\let\auto@bib@innerbib\@empty
\bibitem [{\citenamefont {Cardy}\ and\ \citenamefont
  {Ostlund}(1982)}]{Cardy+82}%
  \BibitemOpen
  \bibfield  {author} {\bibinfo {author} {\bibfnamefont {J.~L.}\ \bibnamefont
  {Cardy}}\ and\ \bibinfo {author} {\bibfnamefont {S.}~\bibnamefont
  {Ostlund}},\ }\href@noop {} {\bibfield  {journal} {\bibinfo  {journal} {Phys.
  Rev. B}\ }\textbf {\bibinfo {volume} {25}},\ \bibinfo {pages} {6899}
  (\bibinfo {year} {1982})}\BibitemShut {NoStop}%
\bibitem [{\citenamefont {Toner}\ and\ \citenamefont
  {DiVincenzo}(1990)}]{Toner+90}%
  \BibitemOpen
  \bibfield  {author} {\bibinfo {author} {\bibfnamefont {J.}~\bibnamefont
  {Toner}}\ and\ \bibinfo {author} {\bibfnamefont {D.~P.}\ \bibnamefont
  {DiVincenzo}},\ }\href@noop {} {\bibfield  {journal} {\bibinfo  {journal}
  {Phys. Rev. B}\ }\textbf {\bibinfo {volume} {41}},\ \bibinfo {pages} {632}
  (\bibinfo {year} {1990})}\BibitemShut {NoStop}%
\bibitem [{\citenamefont {Hwa}\ and\ \citenamefont {Fisher}(1994)}]{Hwa+94}%
  \BibitemOpen
  \bibfield  {author} {\bibinfo {author} {\bibfnamefont {T.}~\bibnamefont
  {Hwa}}\ and\ \bibinfo {author} {\bibfnamefont {D.~S.}\ \bibnamefont
  {Fisher}},\ }\href@noop {} {\bibfield  {journal} {\bibinfo  {journal} {Phys.
  Rev. Lett.}\ }\textbf {\bibinfo {volume} {72}},\ \bibinfo {pages} {2466}
  (\bibinfo {year} {1994})}\BibitemShut {NoStop}%
\bibitem [{\citenamefont {Carpentier}\ and\ \citenamefont
  {Le~Doussal}(1997)}]{Carpentier+97}%
  \BibitemOpen
  \bibfield  {author} {\bibinfo {author} {\bibfnamefont {D.}~\bibnamefont
  {Carpentier}}\ and\ \bibinfo {author} {\bibfnamefont {P.}~\bibnamefont
  {Le~Doussal}},\ }\href@noop {} {\bibfield  {journal} {\bibinfo  {journal}
  {Phys. Rev. B}\ }\textbf {\bibinfo {volume} {55}},\ \bibinfo {pages} {12128}
  (\bibinfo {year} {1997})}\BibitemShut {NoStop}%
\bibitem [{\citenamefont {Giamarchi}\ and\ \citenamefont
  {Le~Doussal}(1997)}]{GiamarchiLeDoussalBookYoung}%
  \BibitemOpen
  \bibfield  {author} {\bibinfo {author} {\bibfnamefont {T.}~\bibnamefont
  {Giamarchi}}\ and\ \bibinfo {author} {\bibfnamefont {P.}~\bibnamefont
  {Le~Doussal}},\ }in\ \href@noop {} {\emph {\bibinfo {booktitle} {Spin glasses
  and random fields}}},\ \bibinfo {editor} {edited by\ \bibinfo {editor}
  {\bibfnamefont {A.}~\bibnamefont {Young}}}\ (\bibinfo {address} {Singapore},\
  \bibinfo {year} {1997})\BibitemShut {NoStop}%
\bibitem [{\citenamefont {Fisher}(1989)}]{fisher89vortex}%
  \BibitemOpen
  \bibfield  {author} {\bibinfo {author} {\bibfnamefont {M.~P.~A.}\
  \bibnamefont {Fisher}},\ }\href@noop {} {\bibfield  {journal} {\bibinfo
  {journal} {Phys. Rev. Lett.}\ }\textbf {\bibinfo {volume} {62}},\ \bibinfo
  {pages} {1415} (\bibinfo {year} {1989})}\BibitemShut {NoStop}%
\bibitem [{\citenamefont {Blatter}\ \emph {et~al.}(1994)\citenamefont
  {Blatter}, \citenamefont {Feigel'man}, \citenamefont {Geshkenbein},
  \citenamefont {Larkin},\ and\ \citenamefont {Vinokur}}]{Blatter+94}%
  \BibitemOpen
  \bibfield  {author} {\bibinfo {author} {\bibfnamefont {G.}~\bibnamefont
  {Blatter}}, \bibinfo {author} {\bibfnamefont {M.~V.}\ \bibnamefont
  {Feigel'man}}, \bibinfo {author} {\bibfnamefont {V.~B.}\ \bibnamefont
  {Geshkenbein}}, \bibinfo {author} {\bibfnamefont {A.~I.}\ \bibnamefont
  {Larkin}}, \ and\ \bibinfo {author} {\bibfnamefont {V.~M.}\ \bibnamefont
  {Vinokur}},\ }\href@noop {} {\bibfield  {journal} {\bibinfo  {journal} {Rev.
  Mod. Phys.}\ }\textbf {\bibinfo {volume} {66}},\ \bibinfo {pages} {1125}
  (\bibinfo {year} {1994})}\BibitemShut {NoStop}%
\bibitem [{\citenamefont {Nattermann}\ and\ \citenamefont
  {Scheidl}(2000)}]{Nattermann+00}%
  \BibitemOpen
  \bibfield  {author} {\bibinfo {author} {\bibfnamefont {T.}~\bibnamefont
  {Nattermann}}\ and\ \bibinfo {author} {\bibfnamefont {S.}~\bibnamefont
  {Scheidl}},\ }\href@noop {} {\bibfield  {journal} {\bibinfo  {journal} {Adv.
  Phys.}\ }\textbf {\bibinfo {volume} {49}},\ \bibinfo {pages} {607} (\bibinfo
  {year} {2000})}\BibitemShut {NoStop}%
\bibitem [{\citenamefont {Feldman}(2001)}]{Feldman2001}%
  \BibitemOpen
  \bibfield  {author} {\bibinfo {author} {\bibfnamefont {D.}~\bibnamefont
  {Feldman}},\ }\href@noop {} {\bibfield  {journal} {\bibinfo  {journal} {Int.
  J. Mod. Phys. B}\ }\textbf {\bibinfo {volume} {15}},\ \bibinfo {pages} {2945}
  (\bibinfo {year} {2001})}\BibitemShut {NoStop}%
\bibitem [{\citenamefont {Tissier}\ and\ \citenamefont
  {Tarjus}(2006)}]{TarjusTissier2005}%
  \BibitemOpen
  \bibfield  {author} {\bibinfo {author} {\bibfnamefont {M.}~\bibnamefont
  {Tissier}}\ and\ \bibinfo {author} {\bibfnamefont {G.}~\bibnamefont
  {Tarjus}},\ }\href@noop {} {\bibfield  {journal} {\bibinfo  {journal} {Phys.
  Rev. Lett.}\ }\textbf {\bibinfo {volume} {96}},\ \bibinfo {pages} {087202}
  (\bibinfo {year} {2006})}\BibitemShut {NoStop}%
\bibitem [{\citenamefont {Fedorenko}\ and\ \citenamefont
  {K\"uhnel}(2007)}]{FEDORENKO}%
  \BibitemOpen
  \bibfield  {author} {\bibinfo {author} {\bibfnamefont {A.~A.}\ \bibnamefont
  {Fedorenko}}\ and\ \bibinfo {author} {\bibfnamefont {F.}~\bibnamefont
  {K\"uhnel}},\ }\href@noop {} {\bibfield  {journal} {\bibinfo  {journal}
  {Phys. Rev. B}\ }\textbf {\bibinfo {volume} {75}},\ \bibinfo {pages} {174206}
  (\bibinfo {year} {2007})}\BibitemShut {NoStop}%
\bibitem [{\citenamefont {Radzihovsky}\ and\ \citenamefont
  {Toner}(1999)}]{RADZ1}%
  \BibitemOpen
  \bibfield  {author} {\bibinfo {author} {\bibfnamefont {L.}~\bibnamefont
  {Radzihovsky}}\ and\ \bibinfo {author} {\bibfnamefont {J.}~\bibnamefont
  {Toner}},\ }\href@noop {} {\bibfield  {journal} {\bibinfo  {journal} {Phys.
  Rev. B}\ }\textbf {\bibinfo {volume} {60}},\ \bibinfo {pages} {206} (\bibinfo
  {year} {1999})}\BibitemShut {NoStop}%
\bibitem [{\citenamefont {Giamarchi}\ and\ \citenamefont
  {Le~Doussal}(1995)}]{Giamarchi+95PhysRevB.52.1242}%
  \BibitemOpen
  \bibfield  {author} {\bibinfo {author} {\bibfnamefont {T.}~\bibnamefont
  {Giamarchi}}\ and\ \bibinfo {author} {\bibfnamefont {P.}~\bibnamefont
  {Le~Doussal}},\ }\href@noop {} {\bibfield  {journal} {\bibinfo  {journal}
  {Phys. Rev. B}\ }\textbf {\bibinfo {volume} {52}},\ \bibinfo {pages} {1242}
  (\bibinfo {year} {1995})}\BibitemShut {NoStop}%
\bibitem [{\citenamefont
  {Le~Doussal}(2010{\natexlab{a}})}]{LeDoussal2010Book1}%
  \BibitemOpen
  \bibfield  {author} {\bibinfo {author} {\bibfnamefont {P.}~\bibnamefont
  {Le~Doussal}},\ }\href@noop {} {\bibfield  {journal} {\bibinfo  {journal}
  {Int. J. Mod. Phys.}\ }\textbf {\bibinfo {volume} {24}},\ \bibinfo {pages}
  {3855} (\bibinfo {year} {2010}{\natexlab{a}})}\BibitemShut {NoStop}%
\bibitem [{\citenamefont {Fisher}(1997)}]{DSFISHER}%
  \BibitemOpen
  \bibfield  {author} {\bibinfo {author} {\bibfnamefont {D.~S.}\ \bibnamefont
  {Fisher}},\ }\href@noop {} {\bibfield  {journal} {\bibinfo  {journal} {Phys.
  Rev. Lett.}\ }\textbf {\bibinfo {volume} {78}},\ \bibinfo {pages} {1964}
  (\bibinfo {year} {1997})}\BibitemShut {NoStop}%
\bibitem [{\citenamefont {Klein}\ \emph {et~al.}(2001)\citenamefont {Klein},
  \citenamefont {Joumard}, \citenamefont {Blanchard}, \citenamefont {Marcus},
  \citenamefont {Cubitt}, \citenamefont {Giamarchi},\ and\ \citenamefont
  {Le~Doussal}}]{KleinJoumardBlanchardMarcusCubittGiamarchiLeDoussal2001}%
  \BibitemOpen
  \bibfield  {author} {\bibinfo {author} {\bibfnamefont {T.}~\bibnamefont
  {Klein}}, \bibinfo {author} {\bibfnamefont {I.}~\bibnamefont {Joumard}},
  \bibinfo {author} {\bibfnamefont {S.}~\bibnamefont {Blanchard}}, \bibinfo
  {author} {\bibfnamefont {J.}~\bibnamefont {Marcus}}, \bibinfo {author}
  {\bibfnamefont {R.}~\bibnamefont {Cubitt}}, \bibinfo {author} {\bibfnamefont
  {T.}~\bibnamefont {Giamarchi}}, \ and\ \bibinfo {author} {\bibfnamefont
  {P.}~\bibnamefont {Le~Doussal}},\ }\href@noop {} {\bibfield  {journal}
  {\bibinfo  {journal} {Nature}\ }\textbf {\bibinfo {volume} {413}},\ \bibinfo
  {pages} {404} (\bibinfo {year} {2001})}\BibitemShut {NoStop}%
\bibitem [{\citenamefont {Bogner}\ \emph {et~al.}(2001)\citenamefont {Bogner},
  \citenamefont {Emig},\ and\ \citenamefont {Nattermann}}]{bogner+01}%
  \BibitemOpen
  \bibfield  {author} {\bibinfo {author} {\bibfnamefont {S.}~\bibnamefont
  {Bogner}}, \bibinfo {author} {\bibfnamefont {T.}~\bibnamefont {Emig}}, \ and\
  \bibinfo {author} {\bibfnamefont {T.}~\bibnamefont {Nattermann}},\
  }\href@noop {} {\bibfield  {journal} {\bibinfo  {journal} {Phys. Rev. B}\
  }\textbf {\bibinfo {volume} {63}},\ \bibinfo {pages} {174501} (\bibinfo
  {year} {2001})}\BibitemShut {NoStop}%
\bibitem [{\citenamefont {Fedorenko}\ \emph {et~al.}()\citenamefont
  {Fedorenko}, \citenamefont {Le~Doussal},\ and\ \citenamefont
  {Wiese}}]{FEDO-US}%
  \BibitemOpen
  \bibfield  {author} {\bibinfo {author} {\bibfnamefont {A.}~\bibnamefont
  {Fedorenko}}, \bibinfo {author} {\bibfnamefont {P.}~\bibnamefont
  {Le~Doussal}}, \ and\ \bibinfo {author} {\bibfnamefont {K.}~\bibnamefont
  {Wiese}},\ }\href@noop {} {\bibinfo  {journal} {(unpublished)}\ }\BibitemShut
  {NoStop}%
\bibitem [{\citenamefont {Ristivojevic}\ \emph {et~al.}(2012)\citenamefont
  {Ristivojevic}, \citenamefont {Le~Doussal},\ and\ \citenamefont
  {Wiese}}]{Ristivojevic+12}%
  \BibitemOpen
\bibfield  {journal} {  }\bibfield  {author} {\bibinfo {author} {\bibfnamefont
  {Z.}~\bibnamefont {Ristivojevic}}, \bibinfo {author} {\bibfnamefont
  {P.}~\bibnamefont {Le~Doussal}}, \ and\ \bibinfo {author} {\bibfnamefont
  {K.~J.}\ \bibnamefont {Wiese}},\ }\href@noop {} {\bibfield  {journal}
  {\bibinfo  {journal} {Phys. Rev. B}\ }\textbf {\bibinfo {volume} {86}},\
  \bibinfo {pages} {054201} (\bibinfo {year} {2012})}\BibitemShut {NoStop}%
\bibitem [{\citenamefont {Perret}\ \emph {et~al.}(2012)\citenamefont {Perret},
  \citenamefont {Ristivojevic}, \citenamefont {Le~Doussal}, \citenamefont
  {Schehr},\ and\ \citenamefont {Wiese}}]{perret+12PhysRevLett.109.157205}%
  \BibitemOpen
  \bibfield  {author} {\bibinfo {author} {\bibfnamefont {A.}~\bibnamefont
  {Perret}}, \bibinfo {author} {\bibfnamefont {Z.}~\bibnamefont
  {Ristivojevic}}, \bibinfo {author} {\bibfnamefont {P.}~\bibnamefont
  {Le~Doussal}}, \bibinfo {author} {\bibfnamefont {G.}~\bibnamefont {Schehr}},
  \ and\ \bibinfo {author} {\bibfnamefont {K.~J.}\ \bibnamefont {Wiese}},\
  }\href@noop {} {\bibfield  {journal} {\bibinfo  {journal} {Phys. Rev. Lett.}\
  }\textbf {\bibinfo {volume} {109}},\ \bibinfo {pages} {157205} (\bibinfo
  {year} {2012})}\BibitemShut {NoStop}%
\bibitem [{\citenamefont {Zhang}\ and\ \citenamefont
  {Radzihovsky}(2013)}]{RADZ2}%
  \BibitemOpen
  \bibfield  {author} {\bibinfo {author} {\bibfnamefont {Q.}~\bibnamefont
  {Zhang}}\ and\ \bibinfo {author} {\bibfnamefont {L.}~\bibnamefont
  {Radzihovsky}},\ }\href@noop {} {\bibfield  {journal} {\bibinfo  {journal}
  {Phys. Rev. E}\ }\textbf {\bibinfo {volume} {87}},\ \bibinfo {pages} {022509}
  (\bibinfo {year} {2013})}\BibitemShut {NoStop}%
\bibitem [{\citenamefont {Zhang}\ and\ \citenamefont
  {Radzihovsky}(2012)}]{RADZ21}%
  \BibitemOpen
  \bibfield  {author} {\bibinfo {author} {\bibfnamefont {Q.}~\bibnamefont
  {Zhang}}\ and\ \bibinfo {author} {\bibfnamefont {L.}~\bibnamefont
  {Radzihovsky}},\ }\href@noop {} {\bibfield  {journal} {\bibinfo  {journal}
  {Europhys. Lett.}\ }\textbf {\bibinfo {volume} {98}},\ \bibinfo {pages}
  {56007} (\bibinfo {year} {2012})}\BibitemShut {NoStop}%
\bibitem [{\citenamefont {Chaikin}\ and\ \citenamefont
  {Lubensky}(1995)}]{Chaikin+}%
  \BibitemOpen
  \bibfield  {author} {\bibinfo {author} {\bibfnamefont {P.~M.}\ \bibnamefont
  {Chaikin}}\ and\ \bibinfo {author} {\bibfnamefont {T.~M.}\ \bibnamefont
  {Lubensky}},\ }\href@noop {} {\emph {\bibinfo {title} {Principles of
  condensed matter physics}}}\ (\bibinfo  {publisher} {Cambridge University
  Press},\ \bibinfo {year} {1995})\BibitemShut {NoStop}%
\bibitem [{\citenamefont {Nozi\`{e}res}(1992)}]{Nozieres}%
  \BibitemOpen
  \bibfield  {author} {\bibinfo {author} {\bibfnamefont {P.}~\bibnamefont
  {Nozi\`{e}res}},\ }in\ \href@noop {} {\emph {\bibinfo {booktitle} {Solids Far
  From Equilibrium}}},\ \bibinfo {editor} {edited by\ \bibinfo {editor}
  {\bibfnamefont {C.}~\bibnamefont {Godr\`{e}che}}}\ (\bibinfo  {publisher}
  {Cambridge University Press, Cambridge},\ \bibinfo {year} {1992})\BibitemShut
  {NoStop}%
\bibitem [{Note1()}]{Note1}%
  \BibitemOpen
  \bibinfo {note} {Note that since it involves order $2m$ in perturbation
  theory in the disorder, i.e., $A_{2m} \sim g^{2m}$, the result (\ref {toner})
  holds only in the limit of $r \to \infty $ at fixed $q$. For a large but
  fixed $r$, it will be more difficult to measure the behavior for larger
  $q$.}\BibitemShut {Stop}%
\bibitem [{\citenamefont {Bauer}\ and\ \citenamefont
  {Bernard}(1996)}]{bauer+96}%
  \BibitemOpen
  \bibfield  {author} {\bibinfo {author} {\bibfnamefont {M.}~\bibnamefont
  {Bauer}}\ and\ \bibinfo {author} {\bibfnamefont {D.}~\bibnamefont
  {Bernard}},\ }\href@noop {} {\bibfield  {journal} {\bibinfo  {journal}
  {Europhys. Lett.}\ }\textbf {\bibinfo {volume} {33}},\ \bibinfo {pages} {255}
  (\bibinfo {year} {1996})}\BibitemShut {NoStop}%
\bibitem [{\citenamefont {Le~Doussal}\ \emph {et~al.}()\citenamefont
  {Le~Doussal}, \citenamefont {Wiese},\ and\ \citenamefont
  {Ristivojevic}}]{led+unpub}%
  \BibitemOpen
  \bibfield  {author} {\bibinfo {author} {\bibfnamefont {P.}~\bibnamefont
  {Le~Doussal}}, \bibinfo {author} {\bibfnamefont {K.}~\bibnamefont {Wiese}}, \
  and\ \bibinfo {author} {\bibfnamefont {Z.}~\bibnamefont {Ristivojevic}},\
  }\href@noop {} {\bibinfo  {journal} {(unpublished)}\ }\BibitemShut {NoStop}%
\bibitem [{\citenamefont {Guruswamy}\ \emph {et~al.}(2000)\citenamefont
  {Guruswamy}, \citenamefont {LeClair},\ and\ \citenamefont
  {Ludwig}}]{Guruswamy+00}%
  \BibitemOpen
\bibfield  {journal} {  }\bibfield  {author} {\bibinfo {author} {\bibfnamefont
  {S.}~\bibnamefont {Guruswamy}}, \bibinfo {author} {\bibfnamefont
  {A.}~\bibnamefont {LeClair}}, \ and\ \bibinfo {author} {\bibfnamefont
  {A.~W.~W.}\ \bibnamefont {Ludwig}},\ }\href@noop {} {\bibfield  {journal}
  {\bibinfo  {journal} {Nucl. Phys. B}\ }\textbf {\bibinfo {volume} {583}},\
  \bibinfo {pages} {475} (\bibinfo {year} {2000})}\BibitemShut {NoStop}%
\bibitem [{\citenamefont {Le~Doussal}\ and\ \citenamefont
  {Schehr}(2007)}]{LeDoussal+07}%
  \BibitemOpen
  \bibfield  {author} {\bibinfo {author} {\bibfnamefont {P.}~\bibnamefont
  {Le~Doussal}}\ and\ \bibinfo {author} {\bibfnamefont {G.}~\bibnamefont
  {Schehr}},\ }\href@noop {} {\bibfield  {journal} {\bibinfo  {journal} {Phys.
  Rev. B}\ }\textbf {\bibinfo {volume} {75}},\ \bibinfo {pages} {184401}
  (\bibinfo {year} {2007})}\BibitemShut {NoStop}%
\bibitem [{Note2()}]{Note2}%
  \BibitemOpen
  \bibinfo {note} {A real crystal has a finite correlation length $\xi _B$ for
  the bulk translational order. Here we assume delta correlated disorder and
  therefore capture physics at length scales larger than $\xi _B$.}\BibitemShut
  {Stop}%
\bibitem [{\citenamefont {Giamarchi}(2003)}]{Giamarchi}%
  \BibitemOpen
  \bibfield  {author} {\bibinfo {author} {\bibfnamefont {T.}~\bibnamefont
  {Giamarchi}},\ }\href@noop {} {\emph {\bibinfo {title} {Quantum Physics in
  One Dimension}}}\ (\bibinfo  {publisher} {Clarendon press, Oxford},\ \bibinfo
  {year} {2003})\BibitemShut {NoStop}%
\bibitem [{\citenamefont {Schulz}\ \emph {et~al.}(1988)\citenamefont {Schulz},
  \citenamefont {Villain}, \citenamefont {Br\'{e}zin},\ and\ \citenamefont
  {Orland}}]{Schulz+88}%
  \BibitemOpen
  \bibfield  {author} {\bibinfo {author} {\bibfnamefont {U.}~\bibnamefont
  {Schulz}}, \bibinfo {author} {\bibfnamefont {J.}~\bibnamefont {Villain}},
  \bibinfo {author} {\bibfnamefont {E.}~\bibnamefont {Br\'{e}zin}}, \ and\
  \bibinfo {author} {\bibfnamefont {H.}~\bibnamefont {Orland}},\ }\href@noop {}
  {\bibfield  {journal} {\bibinfo  {journal} {J. Stat. Phys}\ }\textbf
  {\bibinfo {volume} {51}},\ \bibinfo {pages} {1} (\bibinfo {year}
  {1988})}\BibitemShut {NoStop}%
\bibitem [{\citenamefont {Le~Doussal}(2010{\natexlab{b}})}]{LeDoussal2010}%
  \BibitemOpen
  \bibfield  {author} {\bibinfo {author} {\bibfnamefont {P.}~\bibnamefont
  {Le~Doussal}},\ }\href@noop {} {\bibfield  {journal} {\bibinfo  {journal}
  {Ann. Phys.}\ }\textbf {\bibinfo {volume} {325}},\ \bibinfo {pages} {49}
  (\bibinfo {year} {2010}{\natexlab{b}})}\BibitemShut {NoStop}%
\bibitem [{\citenamefont {Zinn-Justin}(2002)}]{Zinn-Justin}%
  \BibitemOpen
  \bibfield  {author} {\bibinfo {author} {\bibfnamefont {J.}~\bibnamefont
  {Zinn-Justin}},\ }\href@noop {} {\emph {\bibinfo {title} {Quantum Field
  Theory and Critical Phenomena}}}\ (\bibinfo  {publisher} {Clarendon Press,
  Oxford},\ \bibinfo {year} {2002})\BibitemShut {NoStop}%
\bibitem [{\citenamefont {Le~Doussal}(2006)}]{LeDoussal2006b}%
  \BibitemOpen
  \bibfield  {author} {\bibinfo {author} {\bibfnamefont {P.}~\bibnamefont
  {Le~Doussal}},\ }\href@noop {} {\bibfield  {journal} {\bibinfo  {journal}
  {Europhys. Lett.}\ }\textbf {\bibinfo {volume} {76}},\ \bibinfo {pages} {457}
  (\bibinfo {year} {2006})}\BibitemShut {NoStop}%
\bibitem [{\citenamefont {Dotsenko}\ and\ \citenamefont
  {Fateev}(1984)}]{DotsenkoFateev1984}%
  \BibitemOpen
  \bibfield  {author} {\bibinfo {author} {\bibfnamefont {V.}~\bibnamefont
  {Dotsenko}}\ and\ \bibinfo {author} {\bibfnamefont {L.}~\bibnamefont
  {Fateev}},\ }\href@noop {} {\bibfield  {journal} {\bibinfo  {journal} {Nucl.
  Phys. B}\ }\textbf {\bibinfo {volume} {240}},\ \bibinfo {pages} {312}
  (\bibinfo {year} {1984})}\BibitemShut {NoStop}%
\bibitem [{\citenamefont {Dotsenko}\ and\ \citenamefont
  {Fateev}(1985)}]{DotsenkoFateev1985}%
  \BibitemOpen
  \bibfield  {author} {\bibinfo {author} {\bibfnamefont {V.}~\bibnamefont
  {Dotsenko}}\ and\ \bibinfo {author} {\bibfnamefont {L.}~\bibnamefont
  {Fateev}},\ }\href@noop {} {\bibfield  {journal} {\bibinfo  {journal} {Nucl.
  Phys. B}\ }\textbf {\bibinfo {volume} {251}},\ \bibinfo {pages} {691}
  (\bibinfo {year} {1985})}\BibitemShut {NoStop}%
\bibitem [{\citenamefont {Dotsenko}\ \emph {et~al.}(1995)\citenamefont
  {Dotsenko}, \citenamefont {Picco},\ and\ \citenamefont
  {Pujol}}]{Dotsenko+95}%
  \BibitemOpen
  \bibfield  {author} {\bibinfo {author} {\bibfnamefont {V.~S.}\ \bibnamefont
  {Dotsenko}}, \bibinfo {author} {\bibfnamefont {M.}~\bibnamefont {Picco}}, \
  and\ \bibinfo {author} {\bibfnamefont {P.}~\bibnamefont {Pujol}},\
  }\href@noop {} {\bibfield  {journal} {\bibinfo  {journal} {Nucl. Phys. B}\
  }\textbf {\bibinfo {volume} {455}},\ \bibinfo {pages} {701} (\bibinfo {year}
  {1995})}\BibitemShut {NoStop}%
\bibitem [{\citenamefont {Guida}\ and\ \citenamefont
  {Magnoli}(1998)}]{Guida+98}%
  \BibitemOpen
  \bibfield  {author} {\bibinfo {author} {\bibfnamefont {R.}~\bibnamefont
  {Guida}}\ and\ \bibinfo {author} {\bibfnamefont {N.}~\bibnamefont
  {Magnoli}},\ }\href@noop {} {\bibfield  {journal} {\bibinfo  {journal} {Int.
  J. Mod. Phys. A}\ }\textbf {\bibinfo {volume} {13}},\ \bibinfo {pages} {1145}
  (\bibinfo {year} {1998})}\BibitemShut {NoStop}%
\bibitem [{\citenamefont {Prudnikov}\ \emph {et~al.}(1998)\citenamefont
  {Prudnikov}, \citenamefont {Brychkov},\ and\ \citenamefont
  {Marichev}}]{Prudnikov}%
  \BibitemOpen
  \bibfield  {author} {\bibinfo {author} {\bibfnamefont {A.}~\bibnamefont
  {Prudnikov}}, \bibinfo {author} {\bibfnamefont {Y.}~\bibnamefont {Brychkov}},
  \ and\ \bibinfo {author} {\bibfnamefont {O.}~\bibnamefont {Marichev}},\
  }\href@noop {} {\emph {\bibinfo {title} {Integrals and Series: Volume 3: More
  Special Functions}}}\ (\bibinfo  {publisher} {Gordon and Breach, Amsterdam},\
  \bibinfo {year} {1998})\BibitemShut {NoStop}%
\bibitem [{fun()}]{functions.wolfram.07.27.17.0018.0-spliting}%
  \BibitemOpen
  \href@noop {} {\bibinfo  {journal}
  {http://functions.wolfram.com/07.27.17.0018.01}\ }\BibitemShut {NoStop}%
\bibitem [{\citenamefont {Ahlfors}(1979)}]{Ahlfors}%
  \BibitemOpen
\bibfield  {journal} {  }\bibfield  {author} {\bibinfo {author} {\bibfnamefont
  {L.~V.}\ \bibnamefont {Ahlfors}},\ }\href@noop {} {\emph {\bibinfo {title}
  {Complex Analysis}}}\ (\bibinfo  {publisher} {McGraw-Hill, Inc.},\ \bibinfo
  {year} {1979})\BibitemShut {NoStop}%
\end{thebibliography}

%

\end{document}